\documentclass[cmp]{svjour}  
\usepackage{amsmath}
\usepackage{amsfonts,amssymb,epic,eepic}
\usepackage{graphics}
\usepackage[dvips]{graphicx}
\usepackage[dvips]{hyperref}
\usepackage{epsfig}
\journalname{Communications in Mathematical Physics}

\newcommand{\hr}{{\cal H}}

\newcommand{\C}{{\mathbb C}}
\newcommand{\Fock}{{\{0,1\}^*}}

\newcommand{\N}{{\mathbb N}}
\newcommand{\idn}{\mathbf{1}}
\newcommand{\Z}{{\mathbb Z}}

\newcommand{\eps}{{\varepsilon}}        

\newcommand{\lineclear}{{\rule{0pt}{0pt}\nopagebreak\par\nopagebreak\noindent}}

\begin{document}

\title{{Entropy and Quantum Kolmogorov Complexity: A Quantum Brudno's Theorem}}

\author{Fabio Benatti\inst{1}, Tyll Kr\"uger\inst{2}\fnmsep\inst{3},
Markus M\"uller\inst{2}, Rainer Siegmund-Schultze\inst{2}, Arleta Szko\l a\inst{2}}

\institute{University of Trieste, Department of Theoretical Physics, Strada Costiera,
11, 34014 Trieste, Italy. \email{benatti@ts.infn.it} \and
Technische Universit\"at Berlin, Fakult\"at II - Mathematik und Naturwissenschaften,
Institut f\"ur Mathematik MA 7-2, Stra\ss e des 17. Juni 136, 10623 Berlin, Germany.\\
\email{\{tkrueger, mueller, siegmund, szkola\}@math.tu-berlin.de} \and
Universit\"at Bielefeld, Fakult\"at f\"ur Mathematik, Universit\"atsstr. 25,
33619 Bielefeld, Germany.
}

\authorrunning{F. Benatti, T. Kr\"uger, M. M\"uller, Ra. Siegmund-Schultze, A. Szko\l a}

\date{Received: 15 June 2005 / Accepted: 6 January 2006}
\communicated{M. B. Ruskai}

\maketitle
\begin{abstract}
In classical information theory, entropy rate and algorithmic complexity
per symbol are related by a theorem of Brudno. In this paper, we
prove a quantum version of this theorem, connecting the von
Neumann entropy rate and two notions of quantum
Kolmogorov complexity, both based on the shortest
qubit descriptions of qubit strings that, run by a universal quantum
Turing machine, reproduce them as outputs.
\end{abstract}

\section{Introduction}
In recent years, the theoretical and experimental use of quantum
systems to store, transmit and process information has spurred the
study of how much of classical information theory can be extended to the
new territory of quantum information and, vice versa, how much novel
strategies and concepts are needed that have no classical
counterpart.

We shall compare the relations between the rate at
which entropy is produced by classical, respectively quantum,
ergodic sources, and the complexity of the emitted strings of bits,
respectively qubits.

According to Kolmogorov \cite{Kolmogorov65}, the complexity of a bit
string is the minimal
length of a program for a Turing machine ({\sl TM}) that produces the string.
More in detail, the algorithmic complexity $K(\mathbf{i}^{(n)})$ of
a string $\mathbf{i}^{(n)}$ is the length (counted in the number of
bits) of the shortest program $p$ that fed into a universal {\sl TM}
({\sl UTM})
$\mathfrak{U}$ yields the string as output, i.e.
$\mathfrak{U}(p)=\mathbf{i}^{(n)}$. For infinite sequences
$\mathbf{i}$, in analogy with the entropy rate, one defines the
\textit{complexity rate} as $k(\mathbf{i})
:=\lim_n\frac{1}{n}K(\mathbf{i}^{(n)})$, where
$\mathbf{i}^{(n)}$ is the string consisting of the first $n$ bits of
$\mathbf{i}$, \cite{AlYak}. The universality of $\mathfrak{U}$ implies
that changing
the {\sl UTM}, the difference in the complexity of a given string is
bounded by a constant independent of the string; it follows that the
complexity rate $k(\mathbf{i})$ is {\sl UTM}-independent.

Different ways to quantify the complexity of qubit strings
have been put forward; in this paper, we shall be concerned with some
which directly generalize the classical definition by  relating
the complexity of qubit strings with their algorithmic
description by means of quantum Turing machines ({\sl QTM}).

For classical ergodic sources, an important theorem, proved by Brudno
\cite{Brudno} and conjectured before by Zvonkin and Levin \cite{Zvonkin},
establishes that the
entropy rate equals the algorithmic complexity per symbol of almost
all emitted bit strings. We shall show that  this essentially
also holds in quantum information theory.

For stationary classical information sources, the most important
parameter is the \textit{entropy rate}
$h(\pi)=\lim_n\frac{1}{n}H(\pi^{(n)})$, where $H(\pi^{(n)})$ is the
Shannon entropy of the ensembles of strings of length $n$ that are
emitted according to the probability distribution $\pi^{(n)}$.
According to the
Shannon-McMillan-Breiman theorem~\cite{Billingsley,CoverThomas},
$h(\pi)$ represents the optimal compression rate at which the
information provided by classical ergodic sources can be
compressed and then retrieved with negligible probability of error
(in the limit of longer and longer strings). Essentially, $nh(\pi)$
is the number of bits that are needed for reliable compression of
bit strings of length $n$.

Intuitively, the less amount of patterns the emitted strings contain, the
harder will be their compression, which is based on the presence of
regularities and on the elimination of redundancies. From this point
of view, the entropy rate measures the randomness of a classical
source by means of its compressibility on the average, but does not
address the randomness of single strings in the first instance. This
latter problem
was approached by Kolmogorov \cite{Kolmogorov65,Kolmogorov68}, 
(and independently and almost at the
same time by Chaitin \cite{Chaitin}, and Solomonoff
\cite{Solomonoff}), 
in terms of the difficulty of their description by means of algorithms executed
by universal Turing machines ({\sl UTM}), see also~\cite{Vitanyibook}.

On the whole, structureless strings offer no catch for writing down
short programs that fed into a computer produce the given strings
as outputs. The intuitive notion of random strings is thus
mathematically characterized by Kolmogorov by the fact that, for
large $n$, the shortest programs that reproduce them cannot do better
than literal transcription~\cite{USS}.

Intuitively, one expects a connection between the randomness of
single strings and the average randomness of ensembles of strings.
In the classical case, this is exactly the content of a theorem of
Brudno~\cite{Brudno,Whi,Keller,Sow} which states that for
ergodic sources, the complexity rate of $\pi$-almost all infinite sequences
$\mathbf{i}$ coincides with the entropy rate, i.e. $k(\textbf{i})=h(\pi)$.

Quantum sources can be thought as black boxes emitting strings of qubits.
The ensembles of emitted strings of length $n$ are described by
a density operator $\rho^{(n)}$ on the Hilbert spaces $(\C^2)^{\otimes n}$,
which replaces the probability distribution $\pi^{(n)}$ from the
classical case.

The simplest quantum sources are of Bernoulli type: they amount
to infinite quantum spin chains described by shift-invariant states 
characterized by local density matrices $\rho^{(n)}$ over $n$ sites 
with a tensor product structure $\rho^{(n)}=\rho^{\otimes n}
:=\bigotimes_{i=1}^n \rho$, where $\rho$ is a density operator on $\C^2$.

However, typical ergodic states of quantum spin-chains have richer
structures that could be used as quantum sources: the local states
$\rho^{(n)}$, not anymore tensor products, would describe emitted
$n-$qubit strings which are correlated density matrices.

Similarly to classical information sources, quantum stationary sources
(shift-invariant chains) are characterized by their entropy rate
$s:=\lim_n\frac{1}{n}S(\rho^{(n)})$, where $S(\rho^{(n)})$ denotes
the von Neumann entropy of the density matrix $\rho^{(n)}$.

The quantum extension of the Shannon-McMillan Theorem was first
obtained in~\cite{Jozsa} for Bernoulli sources, then
a partial assertion was obtained for
the restricted class of completely ergodic sources in~\cite{Hiai},
and finally in~\cite{QSMPaper}, a complete quantum extension
was shown for general ergodic sources. The latter result
is based on the construction of subspaces of
dimension close to $2^{ns}$, being typical for the source, in the
sense that for sufficiently large block length $n$, their corresponding
orthogonal projectors have an expectation value arbitrarily close to  
$1$ with respect to
the state of the quantum source. These typical subspaces have
subsequently been used to construct compression
protocols~\cite{BjelSzk}.

The concept of a universal quantum Turing machine ({\sl
UQTM}) as a precise mathematical model for quantum computation was
first proposed by Deutsch
\cite{Deutsch}.
The detailed construction of {\sl
UQTMs} can be found in~\cite{BernsteinVazirani,ADMDH}:
these machines work analogously to classical {\sl TM}s, that is they consist of
a read/write head, a set of internal control states and input/output
tapes. However, the local transition functions among the machine's
configurations (the programs or quantum algorithms) are given in
terms of probability amplitudes, implying the possibility of linear
superpositions of the machine's configurations. The quantum algorithms
work reversibly. They correspond
to unitary actions of the {\sl UQTM} as a whole. An element of irreversibility
appears only when the output tape
information is extracted by tracing away the other degrees of
freedom of the {\sl UQTM}. This provides linear
superpositions as well as mixtures of the output tape configurations
consisting of the local states $0,1$ and blanks $\#$, which are
elements of the so-called
\textit{computational basis}. The reversibility of the {\sl UQTM}'s 
time evolution
is to be contrasted with recent models of quantum
computation that are based on measurements on large entangled
states, that is on irreversible processes, subsequently performed in
accordance to the outcomes of the previous ones~\cite{Perdrix}. In
this paper we shall be concerned with Bernstein-Vazirani-type {\sl
UQTMs} whose inputs and outputs may be bit or 
qubit strings \cite{BernsteinVazirani}.

Given the theoretical possibility of universal computing machines
working in agreement with the quantum rules, it was a natural step
to extend the problem of algorithmic
descriptions as a complexity measure to the quantum case. Contrary to
the classical case, where different formulations are equivalent,
several inequivalent possibilities are available in the quantum
setting. In the following, we shall use the definitions
in~\cite{Berthiaume} which, roughly speaking, say that the
algorithmic complexity of a qubit string $\rho$ is the
logarithm in base $2$ of the dimension of the smallest Hilbert space
(spanned by computational basis vectors)
containing a quantum state that, once fed into a {\sl UQTM}, makes 
the {\sl UQTM}
compute the output $\rho$ and halt.

In general, quantum states cannot be perfectly distinguished. Thus, it makes
sense to allow some tolerance in the accuracy of the machine's output.
As explained below, there are two natural ways to deal with this,
leading to two (closely related) different complexity notions $QC^{\searrow 0}$
and $QC^\delta$, which correspond to asymptotically vanishing,
respectively small but fixed tolerance.

Both quantum algorithmic complexities $QC^{\searrow 0}$ and
$QC^\delta$ are thus
measured in terms of the length of
\textit{quantum} descriptions of qubit strings, in contrast to another
definition~\cite{Vitanyi} which defines the complexity of a qubit
string as the length of its shortest \textit{classical} description.
A third
definition~\cite{Gacs} is instead based on an extension of the
classical notion of universal probability to that of universal
density matrices. The study of the relations among these proposals is
still in a very preliminary stage. For an approach to quantum complexity
based on the amount of resources (quantum gates)
needed to implement a quantum circuit reproducing a given qubit
string see~\cite{Briegel1,Briegel2}.\footnote{Other considerations
concerning quantum complexity can be found in \cite{Svozil}
and \cite{Segre}.}

The main result of this work is the proof of a weaker form of
Brudno's theorem, connecting the quantum entropy rate $s$ and the
quantum algorithmic complexities $QC^{\searrow 0}$ and $QC^\delta$ of 
pure states emitted by quantum ergodic sources. It will be proved that
there are
sequences of typical subspaces of $(\C^2)^{\otimes n}$, such
that the complexity rates
$\frac{1}{n}QC^{\searrow 0}(q)$ and $\frac 1 n QC^\delta(q)$ of any of
their pure-state projectors $q$ can be made as close to the entropy
rate $s$ as one wants by choosing $n$ large enough, and there are no such sequences with a smaller expected complexity rate.

The paper is divided as follows. In Section 2, a short review
of the $C^*$-algebraic approach to quantum sources is given, while
Section 3 states
as our main result a
quantum version of Brudno's theorem. In Section 4, a detailed survey
of {\sl QTM}s and of the notion of \textit{quantum
Kolmogorov complexity} is presented. In Section 5, based on
a quantum counting argument, a lower bound is given for the
quantum Kolmogorov complexity per qubit, while an upper bound is
obtained in Section 5 by explicit construction of a short quantum
algorithm able to reproduce any pure state projector $q$ belonging
to a particular sequence of high probability subspaces.

\section{Ergodic Quantum Sources}
\label{SecErgodicQuantumSources}

In order to  formulate our main result rigorously, we start with a
brief introduction to the relevant concepts of the formalism of
quasi-local $C^*$-algebras which is the most suited one for dealing with
quantum spin chains. At the same time, we shall fix the notations.

We shall consider the lattice $\mathbb{Z}$ and assign to each site
$x \in \mathbb{Z}$  a $C^*$-algebra $\mathcal{A}_x$ being
a copy of a fixed finite-dimensional algebra $\mathcal{A}$, in the sense
that there exists a $*$-isomorphism $i_x:\mathcal{A}\to\mathcal{A}_x$.
To simplify notations, we write $a\in\mathcal{A}_x$ for 
$i_x(a)\in\mathcal{A}_x$ and $a\in\mathcal{A}$.
The algebra of observables associated to a finite $\Lambda \subset
\mathbb{Z}$ is defined by $\mathcal{A}_{\Lambda}:= \bigotimes _{x
\in \Lambda} \mathcal{A}_x$. Observe that for $\Lambda \subset
\Lambda^{'}$ we have $
\mathcal{A}_{\Lambda^{'}}=\mathcal{A}_{\Lambda}\otimes
\mathcal{A}_{\Lambda^{'} \backslash \Lambda}$ and there is a
canonical embedding of $\mathcal{A}_{\Lambda}$ into
$\mathcal{A}_{\Lambda^{'}}$ given by $a \mapsto a \otimes
\mathbf{1}_{\Lambda^{'}\backslash  \Lambda}$,  where $a \in
\mathcal{A}_{\Lambda}$ and $\mathbf{1}_{\Lambda^{'} \backslash
\Lambda}$ denotes the identity of $\mathcal{A}_{\Lambda^{'}
\backslash \Lambda}$. The infinite-dimensional quasi-local
$C^*$-algebra $\mathcal{A}^{\infty}$ is the norm completion of the
normed algebra $\bigcup_{\Lambda \subset \mathbb{Z}}
\mathcal{A}_{\Lambda}$, where the union is taken over all finite
subsets $\Lambda$.

In the present paper, we mainly deal with qubits, which are the
quantum counterpart of classical bits.
Thus, in the following, we restrict our considerations to the case where
$\mathcal{A}$  is the algebra of observables of a qubit, i.e. the
algebra $\mathcal{M}_2(\mathbb{C})$ of $2\times 2$ matrices acting on
$\mathbb{C}^2$. Since every finite-dimensional unital $C^*$-Algebra $\mathcal{A}$
is $*$-isomorphic to a subalgebra of $\mathcal{M}_2(\C)^{\otimes D}$
for some $D\in\N$, our results contain the general case of arbitrary $\mathcal{A}$.
Moreover, the case of classical bits is covered by $\mathcal{A}$ 
being the subalgebra of $\mathcal{M}_2(\mathbb{C})$ consisting of
diagonal matrices only.

Similarly, we think of  $\mathcal{A}_{\Lambda}$ as the algebra of
observables of qubit strings of length $|\Lambda|$, namely the algebra
$\mathcal{M}_{2^{|\Lambda|}}(\mathbb{C})
=\mathcal{M}_2(\mathbb{C})^{\otimes|\Lambda|}$ of 
$2^{|\Lambda|}\times 2^{|\Lambda|}$ matrices acting on the Hilbert
space $\mathcal{H}_\Lambda:=(\mathbb{C}^2)^{\otimes |\Lambda|}$. The quasi-local
algebra  $\mathcal{A}^{\infty}$ corresponds to the doubly-infinite
qubit strings.

The (right) shift $\tau$ is a $*$-automorphism on $\mathcal{A}^\infty$
uniquely defined by its action on local observables
\begin{eqnarray}
   \tau:a\in\mathcal{A}_{[m,n]}\mapsto a\in\mathcal{A}_{[m+1,n+1]}
\end{eqnarray}
where $[m,n] \subset \mathbb{Z}$ is an integer interval.

A state $\Psi$ on $\mathcal{A}^{\infty}$ is a normalized positive
linear functional on $\mathcal{A}^\infty$. Each local state
$\Psi_{\Lambda}:=\Psi\upharpoonright \mathcal{A}_{\Lambda}$,
$\Lambda \subset \mathbb{Z} $ finite, corresponds to a density
operator $\rho_\Lambda \in \mathcal{A}_{\Lambda}$ by the relation
$\Psi_{\Lambda}(a)= \hbox{Tr}\left(\rho_\Lambda a\right)$, for all
$a \in \mathcal{A}_\Lambda$, where $\rm{Tr}$ is the trace on $(\C^2)^{\otimes |\Lambda|}$.
The density operator $\rho_\Lambda$
is a positive matrix acting on the Hilbert space 
$\mathcal{H}_\Lambda$ associated with
$\mathcal{A}_\Lambda$ satisfying the normalization
condition $\hbox{Tr}\rho_\Lambda=1$. The simplest $\rho_\Lambda$ 
correspond to one-dimensional projectors 
$P:=\vert\psi_\Lambda\rangle\langle\psi_\Lambda\vert$ onto
vectors
$\vert\psi_\Lambda\rangle\in\mathcal{H}_\Lambda$
and are called pure states, while
general density operators are linear convex combinations of
one-dimensional projectors: $\rho_\Lambda=\sum_i\lambda_i
\vert\psi^i_\Lambda\rangle\langle\psi^i_\Lambda\vert$, $\lambda_i\geq 0$,
$\sum_j\lambda_j=1$.
We denote by $\mathcal{T}^+_1(\mathcal{H})$
the convex set of density operators acting on a 
(possibly infinite-dimensional) Hilbert space $\mathcal{H}$, whence
$\rho_\Lambda\in\mathcal{T}^+_1(\mathcal{H}_\Lambda)$.

A state $\Psi $ on
$\mathcal{A}^{\infty}$ corresponds one-to-one to a family of
density operators $\rho_\Lambda \in \mathcal{A}_{\Lambda}$, $\Lambda
\subset \mathbb{Z} $ finite, fulfilling the consistency condition
$\rho_{\Lambda}= \hbox{Tr}_{\Lambda ' \backslash
\Lambda}\left(\rho_{\Lambda '}\right)$ for $\Lambda\subset \Lambda'$,
where $\hbox{Tr}_{\Lambda}$
denotes the partial trace over the local algebra
$\mathcal{A}_{\Lambda}$ which is computed with respect to any
orthonormal basis in the associated Hilbert space 
$\mathcal{H}_\Lambda$. 
Notice that a state $\Psi$ with $\Psi \circ
T= \Psi$, i.e. a shift-invariant state, is uniquely determined by a
consistent sequence of density operators
$\rho^{(n)}:=\rho_{\Lambda(n)}$ in 
$\mathcal{A}^{(n)}:= \mathcal{A}_{\Lambda(n)}$
corresponding to the local states $\Psi^{(n)}:= \Psi_{\Lambda(n)}$,
where  $\Lambda(n)$ denotes the integer interval $[1,n]\subset
\mathbb{Z}$, for each  $n \in \mathbb{N}$.

As motivated in the introduction, in the information-theoretical
context, we interpret the tuple $(\mathcal{A}^{\infty}, \Psi)$
describing the quantum spin chain as a stationary quantum source.

The von Neumann entropy of a density matrix $\rho$ is 
$S(\rho ):=-\hbox{Tr}(\rho \log \rho)$. By the subadditivity of $S$ for a
shift-invariant state $\Psi$ on $\mathcal{A}^{\infty}$, the
following limit, the quantum entropy rate, exists $$
s(\Psi):=\lim_{n\to\infty}\frac{1}{n}S(\rho^{(n)})\ .
$$
The set of shift-invariant states  on $\mathcal{A}^{\infty}$ is
convex and compact in the  weak$*$-topology. The extremal points of
this set are called ergodic states: they are those states which cannot
be decomposed into linear convex combinations of other shift-invariant states.
Notice that in particular the
shift-invariant product states defined by a sequence of density
matrices $\rho^{(n)}= \rho^{\otimes n}$, $n \in \N$, where $\rho $ is a fixed
$2 \times 2$ density matrix, are ergodic. They are the quantum
counterparts of Bernoulli (i.i.d.) processes. Most of the results in
quantum information theory concern such sources, but, as mentioned
in the introduction, more general ergodic quantum sources allowing 
correlations  can be considered.

More concretely, the typical quantum source that has
first been considered was a finite-dimensional quantum system
emitting vector states $|v_i\rangle \in \mathbb{C}^2$ with
probabilities $p(i)$. The
state of such a source is the density matrix
$\rho=\sum_i p(i) \vert v_i\rangle\langle v_i\vert$ being an element of
the full matrix algebra $\mathcal{M}_2(\C)$; furthermore, the
most natural source of qubit strings of length $n$ is the one that emits
vectors $|v_i\rangle$ independently one after the other at each stroke of
time.\footnote{Here we use Dirac's bra-ket notation, where a bra 
$\vert v\rangle$ is a vector in a Hilbert space and a ket $\langle
v\vert$ is its dual vector.} The corresponding state after $n$
emissions is thus the tensor product
\begin{eqnarray*}
\rho^{\otimes n}&=&
\sum_{i_1i_2\cdots i_n}p(i_1)p(i_2)\cdots
p(i_n)\vert v_{i_1}\otimes v_{i_2}\otimes\cdots
v_{i_n}\rangle \langle v_{i_1}\otimes v_{i_2}\otimes\cdots
v_{i_n}\vert\ .
\end{eqnarray*}

In the following, we shall deal with the more general case of {\em ergodic}
sources defined above, which naturally appear e.g. in statistical
mechanics (compare 1D spin chains with finite-range interaction).

When restricted to act only on $n$ successive chain sites, namely on
the local algebra $\mathcal{A}^{(n)}=\mathcal{M}_{2^n}(\mathbb{C})$,
these states correspond to density matrices
$\rho^{(n)}$ acting on $(\C^2)^{\otimes n}$ which are not simply
tensor products, but may contain classical correlations and
entanglement. 
The qubit strings of length $n$ emitted by these sources are generic
density matrices $\sigma$ acting on $(\C^2)^{\otimes n}$, 
which are compatible with the state of the source $\Psi$ in the sense
that $\textrm{supp } \sigma \leq \textrm{supp }\rho^{(n)}$, where 
$\textrm{supp }\sigma$ denotes the support projector of the operator 
$\sigma$, that is the orthogonal projection onto the subspace where
$\sigma$ cannot vanish.
More concretely, $\rho^{(n)}$ can be decomposed in uncountably many
different ways into convex decompositions 
$\rho^{(n)}=\sum_i\lambda_i\sigma^{(n)}_i$ in terms of other density
matrices $\sigma^{(n)}_i$ on the local algebra $\mathcal{A}^{(n)}$ each
one of which describes a possible qubit string of length $n$
emitted by the source.

\section{Main Theorem}
It turns out that the rates of the complexities
$QC^{\searrow 0}$ (approximation-scheme complexity) and
$QC^\delta$ (finite-accuracy complexity)
of the typical pure states of qubit strings generated by an
ergodic quantum source $(\mathcal{A}^\infty, \Psi)$ are
asymptotically equal to the entropy rate $s(\Psi)$ of the
source. A precise formulation of this result is the content of the 
following theorem. It can be seen  as a quantum extension of Brudno's
theorem as a convergence in probability
statement, while the original formulation of Brudno's result is an
almost sure statement.

We remark that a proper introduction to the
concept of quantum Kolmogorov complexity needs
some further considerations. We postpone this task to the next
section.\\
In the remainder of this paper, we call a sequence of projectors 
$p_n \in \mathcal{A}^{(n)}$, $n \in \N$, satisfying 
$\lim_{n \to \infty} \Psi^{(n)}(p_n)=1$
a {\sl sequence of $\Psi$-typical projectors}.

\begin{theorem}[Quantum Brudno Theorem]
\label{TheQBrudno}
\lineclear
Let $(\mathcal{A}^\infty,\Psi)$ be an ergodic quantum source with 
entropy rate $s$.
For every $\delta>0$, there exists a sequence of
$\Psi$-typical projectors $q_n(\delta)\in\mathcal{A}^{(n)}$, 
$n \in \N$, i.e. $\lim_{n \to \infty} \Psi^{(n)}(q_n(\delta))=1$, 
such that for $n$ large enough every
one-dimensional projector $q\leq q_n(\delta)$ satisfies
\begin{eqnarray}
& & \frac 1 n QC^{\searrow 0}(q) \in \left(s-\delta,s+\delta\right),\label{eq1}\\
& & \frac 1 n QC^\delta(q)\in\left( s-\delta(2+\delta)s,s+\delta\right).\label{eq2}
\end{eqnarray}
Moreover, $s$ is the optimal expected asymptotic complexity rate, in the sense that every sequence of projectors $q_n\in\mathcal{A}^{(n)}$, 
$n \in \N$, that for large $n$ may be represented as a sum of mutually orthogonal one-dimensional projectors that all violate the lower bounds in (\ref{eq1}) and (\ref{eq2}) for some $\delta>0$,  has an asymptotically vanishing expectation value with respect to $\Psi$.
\end{theorem}

\section{{\sl QTM}s and Quantum Kolmogorov Complexity}

Algorithmic complexity measures the degree of randomness of a single object.
It is defined as the minimal description length of the object,
relative to a certain ''machine'' (classically a {\sl UTM}). In
order to properly introduce a quantum counterpart of Kolmogorov complexity,
we thus have to specify what kind of objects we want to describe (outputs),
what the descriptions (inputs) are made of, and what kind of machines
run the algorithms.

In accordance to the
introduction, we stipulate that inputs and outputs are so-called (pure or mixed)
{\em variable-length qubit strings}, while the reference machines
will be {\sl QTM}s  as defined by Bernstein
and Vazirani \cite{BernsteinVazirani}, in particular universal {\sl
QTM}s.

\subsection{Variable-Length Qubit Strings}
Let $\hr_k:=\left(\C^{\{0,1\}}\right)^{\otimes k}$ be the Hilbert space
of $k$ Qubits ($k\in\N_0$). We write $\C^{\{0,1\}}$ for $\C^2$ to
indicate that we fix two orthonormal {\em computational basis vectors}
$|0\rangle$ and $|1\rangle$. Since we want to allow superpositions of
different lengths $k$, we consider the Hilbert space $\hr_{\Fock}$
defined as
\[
   \hr_{\Fock}:=\bigoplus_{k=0}^\infty \hr_k\,\,.
\]
The classical finite binary strings $\{0,1\}^*$ are identified with
the computational basis
vectors in $\hr_\Fock$, i.e. $\hr_{\Fock}\simeq \ell^2(\{\lambda,
0,1,00,01,\ldots\})$, where $\lambda$ denotes the empty string.
We also use the notation
\[
   \hr_{\leq n}:=\bigoplus_{k=0}^n \hr_k
\]
and treat it as a subspace of $\hr_\Fock$.
A (variable-length) {\em qubit string} $\sigma\in\mathcal{T}_1^+(\hr_\Fock)$
is a density operator on $\hr_\Fock$. We define the {\em length} $\ell(\sigma)\in\N_0\cup\{\infty\}$
of a qubit string $\sigma\in\mathcal{T}_1^+(\hr_\Fock)$ as
\begin{equation}
   \ell(\sigma):=\min\{n\in\N_0\enspace|\enspace
   \sigma\in\mathcal{T}_1^+(\hr_{\leq n})\}
   \label{DefOfLength}
\end{equation}
or as $\ell(\sigma)=\infty$ if this set is empty (this will never
occur in the following).

There are two reasons for 
considering variable-length and also mixed qubit strings. First, we want our result
to be as general as possible. Second, a {\sl QTM} will naturally produce
superpositions of qubit strings of different lengths; mixed outputs appear naturally
while tracing out the other parts of the {\sl QTM} (input tape,
control, head) after halting.

In contrast to the classical situation, there are uncountably
many qubit strings that cannot be perfectly distinguished by means
of any quantum measurement. If $\rho,\sigma\in\mathcal{T}_1^+(\hr_\Fock)$ are
two qubit strings with finite length, then we can quantify their distance
in terms of the trace distance
\begin{equation}
   \|\rho-\sigma\|_{\rm Tr}:=\frac 1 2 {\rm Tr} \left\vert \rho-\sigma
   \right\vert=\frac 1 2 \sum_i |\lambda_i |\,\,,
   \label{tr-dist}
\end{equation}
where the $\lambda_i$ are the eigenvalues of the Hermitian operator $|\rho-\sigma|
:=\sqrt{(\rho-\sigma)^*\,(\rho-\sigma)}$.

In Subsection~\ref{SubQAC}, we will define Quantum Kolmogorov Complexity $QC(\rho)$
for qubit strings $\rho$. Due to the considerations above,
it cannot be expected that the qubit strings $\rho$ are reproduced
exactly, but it rather makes sense to demand the strings to be generated
within some trace distance $\delta$. Another possibility is
to consider ''approximation schemes'', i.e. to have some parameter $k\in\N$,
and to demand the machine to approximate the desired state better and better
the larger $k$ gets. We will pursue both approaches, corresponding to equations
(\ref{BerthDelta}) and (\ref{Berth}) below.

Note that we can identify every density operator $\rho\in\mathcal{A}^{(n)}$ on the local
$n$-block algebra with its corresponding qubit string $\tilde\rho
\in\mathcal{T}_1^+(\hr_n)\subset \mathcal{T}_1^+(\hr_\Fock)$
such that $\ell(\tilde\rho)=n$. Similarly, we identify qubit strings $\sigma\in\mathcal{T}_1^+(\hr_\Fock)$
of finite length $\ell$ with the state of the input or
output tape of a {\sl QTM} (see Subsection~\ref{BasicDefQTMSec})
containing the state in the cell interval $[0,\ell-1]$
and vice versa.

\subsection{Mathematical Description of {\sl QTMs}}
\label{BasicDefQTMSec}

Due to the equivalence of various models for quantum computation,
the definition of Quantum Kolmogorov Complexity should be rather insensitive
to the details of the underlying machine. Nevertheless, there are some
details which are relevant for our theorem. Thus, we have to give a thorough
definition of what we mean by a {\sl QTM}.

Bernstein and Vazirani (\cite{BernsteinVazirani}, Def. 3.2.2) define
a quantum Turing machine $M$ as a triplet $(\Sigma,Q,\delta)$, where
$\Sigma$ is a finite alphabet with an identified blank symbol $\#$, and
$Q$ is a finite set of states with an identified initial state $q_0$
and final state $q_f\neq q_0$. The function $\delta:Q\times\Sigma\to\tilde\C^{\Sigma\times Q \times \{L,R\}}$
is called the {\em quantum transition function}.
The symbol $\tilde\C$ denotes the set of complex numbers $\alpha\in\C$ such that there is
a deterministic algorithm that computes the real and imaginary parts of $\alpha$ to within $2^{-n}$
in time polynomial in $n$.

One can think of a {\sl QTM} as consisting of
a two-way infinite tape $\mathbf T$ of cells indexed by $\Z$, a control $\mathbf C$, and a single
''read/write'' head $\mathbf H$ that moves along the tape.
A (classical) configuration is a triplet
$\left((\sigma_i)_{i\in\Z},q,k\right)\in\Sigma^\Z\times Q\times \Z$
such that only a finite number of tape cell contents $\sigma_i$ are non-blank ($q$ and $k$
are the state of the control and the position of the head respectively).
Let $C$ be the set of all configu\-rations, and define the Hilbert space
$\hr_{QTM}:=\ell^2(C)$, which can be written as
$\hr_{QTM}=\hr_{\mathbf{C}}\otimes \hr_{\mathbf{H}}
\otimes \hr_{\mathbf{T}}$.

The transition
function $\delta$ generates a linear operator $U_M$ on $\hr_{QTM}$ describing the
time evolution of the {\sl QTM}.
We identify $\sigma\in\mathcal{T}_1^+(\hr_\Fock)$ with the initial state
of $M$ on input $\sigma$, which is according to the definition in
\cite{BernsteinVazirani} a state on  $\hr_{QTM}$ where $\sigma$ is written on the input track
over the cell interval $[0, l(\sigma)-1]$, the empty state
$\#$ is written on the remaining cells of the input track and on the whole
output track, the control is in the initial state $q_0$ and the head is
in position $0$. Then, the state $M^t(\sigma)$
of $M$ on input $\sigma$ at time $t\in\N_0$ is given by $M^t(\sigma)=\left(U_M\right)^t \sigma
\left(U_M^*\right)^t$. The state of the control at time $t$ is thus given by partial trace
over all the other parts of the machine, that is $M_{\mathbf C}^t(\sigma):={\rm Tr}_{\mathbf{H,T}}\left(
M^t(\sigma)\right)$. In accordance with \cite{BernsteinVazirani}, Def. 3.5.1, we say that the
{\sl QTM} $M$ {\em halts at time $t\in\N_0$ on input $\sigma\in\mathcal{T}_1^+(\hr_\Fock)$},
if and only if
\begin{equation}
   \langle q_f|M_{\rm\bf C}^t(\sigma)|q_f\rangle=1\qquad \mbox{and}\quad
   \langle q_f|M_{\rm\bf C}^{t'}(\sigma)|q_f\rangle=0\quad
   \mbox{for every }t'<t\,\,,
   \label{EqHalting}
\end{equation}
where $q_f\in Q$ is the special state of the control (specified in the
definition of $M$) signalling the halting of the computation. 

Denote
by $\tilde \hr (t) \subset \hr_\Fock$ the set of vector inputs with equal
halting time $t$. Observe that the above definition implies that $\hr (t):= \{c\ |\phi \rangle :\ c \in \C,\ |\phi\rangle \in \tilde \hr (t)\}$ is equal to the linear span of $\tilde \hr(t)$, i.e.    $\hr(t)$ is
a linear subspace of $\hr_\Fock$. Moreover for $t \not= t'$ the corresponding
subspaces $\hr(t)$ and $\hr(t')$ are mutually orthogonal,
because otherwise one could perfectly distinguish non-orthogonal vectors by means
of the halting time. It follows that the subset of
$\mathcal{T}_1^+(\hr_\Fock)$ on which a {\sl QTM M} halts is a union $
\bigcup_{t \in \N} \mathcal{T}_1^+(\hr(t))$.

For our purpose, it is useful to consider a special class of {\sl QTMs}
with the property that their tape $\mathbf T$ consists of two different tracks,
an {\em input track} $\mathbf I$ and an {\em output track} $\mathbf O$. This can be achieved by
having an alphabet which is a Cartesian product of two alphabets,
in our case $\Sigma=\{0,1,\#\}\times \{0,1,\#\}$. Then, the tape Hilbert space
$\hr_{\mathbf T}$ can be written as $\hr_{\mathbf{T}}=\hr_{\mathbf{I}}\otimes
\hr_{\mathbf{O}}$.

\begin{definition}[Quantum Turing Machine ({\sl QTM})]
\lineclear
A partial map $M:\mathcal{T}_1^+(\hr_\Fock)\to\mathcal{T}_1^+(\hr_\Fock)$
will be called a {\sl QTM}, if there is a Bernstein-Vazirani two-track QTM $\tilde M=(\Sigma,Q,\delta)$
(see \cite{BernsteinVazirani}, Def. 3.5.5)
with the following properties:
\begin{itemize}
\item $\Sigma=\{0,1,\#\}\times \{0,1,\#\}$,
\item the corresponding time evolution operator $U_{\tilde M}$ is unitary,
\item if $\tilde M$ halts on input $\sigma$ with a variable-length qubit string
$\rho\in\mathcal{T}_1^+(\hr_\Fock)$ on the output track starting in cell $0$
such that the $i$-th cell is empty for every $i\not\in [0,\ell(\rho)-1]$,
then $M(\sigma)=\rho$; otherwise, $M(\sigma)$ is undefined.
\end{itemize}
\end{definition}

In general, different inputs $\sigma$ have different halting times
$t$ and the corresponding outputs are essentially results of
different unitary transformations given by $U_M^t$. However, as the subset of $\mathcal{T}^+_1(\hr_\Fock)$ on which  
$M$ is defined is of the form $\bigcup_{t \in \N}\mathcal{T}_1^+(\hr(t))$,
the action of the partial map
$M$ on this subset  may be extended to a valid quantum operation\footnote{Recall that
quantum operations are trace-preserving completely positive maps on the trace-class
operators $\mathcal{T}(\mathcal{H})$ on the system Hilbert space $\mathcal{H}$, see \cite{holevo}.} on $\mathcal{T}(\hr_\Fock)$: 
\begin{lemma}[{\sl QTM}s are Quantum Operations]
\label{LemmaQTMsAreOperations}
\lineclear
For every {\sl QTM} $M$  there is a quantum operation
$\mathcal{M}:\mathcal{T}(\hr_\Fock)\to\mathcal{T}(\hr_\Fock)$, 
such that for every $\sigma\in \bigcup_{t \in \N}\mathcal{T}_1^+(\hr(t))$
\[
   M(\sigma)=\mathcal{M}(\sigma).
\]
\end{lemma}
{\bf Proof. } Let $\mathcal{B}_t$ and
$\mathcal{B}_{\perp}$ be an orthonormal basis of $\hr(t)$, $t \in \N$, and the
orthogonal complement of $\bigoplus_{t \in \N}\hr(t)$
within $\hr_\Fock$, respectively. 
We add an ancilla Hilbert space $\hr_{\mathbf{A}}:=\ell^2(\N_0)$ to the {\sl QTM},
and define a linear operator $V_M:\hr_\Fock\to\hr_{QTM}\otimes \hr_{\mathbf{A}}$
by specifying its action on the orthonormal basis vectors $\cup_{t \in \N}\mathcal{B}_t\cup
\mathcal{B}_{\perp}$:
\begin{eqnarray}
V_M |b \rangle :=\left\{
  \begin{array}{rl}
\left(\strut U_M^t | b \rangle\right) \otimes |t\rangle&
 \textrm{ if } 
|b\rangle \in \mathcal{B}_t,
\\ 
|b\rangle \otimes |0\rangle&  \textrm{ if } |b\rangle \in \mathcal{B}_{\perp}.
\label{eqONB}
  \end{array}
\right. 
\end{eqnarray}
Since the right hand side of (\ref{eqONB}) is a set of orthonormal vectors in
$\hr_{QTM}\otimes \hr_{\mathbf{A}}$, the map $V_M$ is a partial isometry.
Thus, the map $\sigma\mapsto V_M \sigma V_M^*$ is trace-preserving, completely
positive (\cite{paulsen}). Its composition with the partial trace, given
by $\mathcal{M}(\sigma):= {\rm Tr}_{\mathbf{C,H,I,A}}(V_M \sigma V_M^{*})$, is a quantum operation.
\qed

\subsection{Quantum Algorithmic Complexity}
\label{SubQAC}
The typical case we want to study is the (approximate) reproduction of a
density matrix $\rho\in \mathcal{T}^+_1( \mathcal{A}^{(n)})$
by a  QTM $M$. This means that there is
a ''quantum program'' $\sigma\in \mathcal{T}^+_1(\hr_{\Fock})$,
such that $M(\sigma)\approx \rho$
in a sense explained below.

We are particularly interested in the case that the program $\sigma$
is shorter than $\rho$ itself, i.e. that $\ell(\sigma)<\ell(\rho)$.
On the whole, the minimum possible length $\ell(\sigma)$ for $\rho$
will be defined as the {\sl quantum algorithmic complexity} of
$\rho$.

As already mentioned, there are at least two natural possible
definitions. The first one is to demand only approximate
reproduction of $\rho$ within some trace distance $\delta$. The
second one is based on  the notion of an approximation scheme.
To define the latter, we have to specify what we mean by supplying
a {\sl QTM} with {\em two} inputs, the qubit string and a parameter:
\begin{definition}[Parameter Encoding]
\label{DefEncoding}
\lineclear
Let $k\in\N$ and $\sigma\in\mathcal{T}_1^+(\hr_{\Fock})$. 
We define an encoding $\mathcal{C}:\N 
\times\mathcal{T}_1^+(\hr_{\Fock})\to \mathcal{T}_1^+(\hr_{\Fock})$
of a pair $(k,\sigma)$ into a single string $\mathcal{C}(k,\sigma)$ by
\[
   \mathcal{C}(k,\sigma):=|\tilde k\rangle\langle\tilde k|\otimes \sigma\,\,,
\]
where $\tilde k$ denotes the (classical) string consisting of 
$\lfloor \log k\rfloor$ $1$'s,
followed by one $0$, followed by the $\lfloor \log k\rfloor +1$ binary
digits of $k$, and $|\tilde k\rangle\langle \tilde k|$ is the
corresponding projector in the computational 
basis\footnote{We use the notations 
$\lfloor x \rfloor=\max\{n\in\N\enspace|\enspace n\leq x\}$ and 
$\lceil x \rceil =\min\{n\in\N\enspace|\enspace n\geq x\}$.}. 
For every {\sl QTM} $M$, we set
\[
   M(k,\sigma):=M(\mathcal{C}(k,\sigma))\,\,.
\]
\end{definition}

Note that
\begin{eqnarray}
\label{encoding_length}
\ell(\mathcal{C}(k,\sigma))=2\lfloor \log k\rfloor +2+\ell(\sigma).
\end{eqnarray}
 
The {\sl QTM} $M$
has to be constructed in such a way that it is able to decode both $k$
and $\sigma$
from $\mathcal{C}(k,\sigma)$, which is an easy classical task.
\begin{definition}[Quantum Algorithmic Complexity]
\label{defQK}
\lineclear
Let $M$ be a {\sl QTM} and $\rho\in\mathcal{T}_1^+(\hr_{\Fock})$ a qubit string.
For every $\delta\geq 0$, we define the {\em finite-accuracy quantum
complexity} $QC_M^\delta(\rho)$ as the minimal length $\ell(\sigma)$
of any quantum program  $\sigma \in \mathcal{T}^{+}_1(\hr_{\Fock})$
such that the corresponding output $M(\sigma)$ has trace distance from $\rho$
smaller than $\delta$,
\begin{equation}
   QC_M^\delta(\rho):=\min\left\{
      \ell(\sigma)\enspace \vert \enspace \|
         \rho-M(\sigma)
      \|_{\rm Tr}\leq \delta
   \right\}\,\,.
   \label{BerthDelta}
\end{equation}
Similarly, we define an {\em approximation-scheme quantum
complexity} $QC_M^{\searrow 0}$ by the minimal length $\ell(\sigma)$
of any density operator $\sigma \in \mathcal{T}^{+}_1(\hr_{\Fock})$,
such that when given $M$ as input
together with any integer $k$, the
output $M(k, \sigma)$ has trace distance from $\rho$
smaller than $1/k$:
\begin{equation}
   QC_M^{\searrow 0}(\rho):=\min\left\{
      \ell(\sigma)\ \left\vert\  \|\rho-M(k, \sigma)\|_{\rm Tr}
\leq \frac 1 k \mbox{ for every }k\in\N
   \right.\right\}\ .
   \label{Berth}
\end{equation}
\end{definition}
Some points are worth stressing in connection with the previous
definition:
\begin{itemize}
\item
This definition is essentially equivalent to the definition given by
Berthiaume et. al. in \cite{Berthiaume}. The only technical
difference is that we found it convenient to use the trace distance
rather than the fidelity.
\item
The {\em same} qubit program $\sigma$ is accompanied by a classical
specification of an integer $k$, which tells the program to what accuracy the
computation of the output state must be accomplished.
\item
If $M$ does not have too restricted functionality (for example, if
$M$ is universal, which is discussed below), a noiseless
transmission channel (implementing the identity transformation)
between the input and output tracks can
always be realized: this corresponds to classical literal transcription,
so that automatically $QC_M^{\delta}(\rho)\leq\ell(\rho)+c_M$ for
some constant $c_M$. Of
course, the key point in classical as well as quantum algorithmic
complexity is that there are sometimes much shorter qubit programs
than literal transcription.
\item
The exact choice of the accuracy specification $\frac 1 k$ is not
important; we can choose any computable function that tends to zero
for $k\to\infty$, and we will always get an equivalent definition
(in the sense of being equal up to some constant).

The same is true for the choice of the encoding $\mathcal{C}$:
As long as $k$ and $\sigma$ can both be computably decoded from 
$\mathcal{C}(k,\sigma)$
and as long as there is no way to extract additional information on the
desired output $\rho$
from the $k$-description part of $\mathcal{C}(k,\sigma)$, the results will
be equivalent up to some constant.
\end{itemize}
Both quantum algorithmic complexities $QC^\delta$ and 
$QC^{\searrow 0}$ are related to each other in a useful way:

\begin{lemma}[Relation between Q-Complexities]
\label{LemRelation} For every {\sl QTM} $M$ and every $k\in\N$, 
we have the relation
\begin{equation}
QC_M^{1/k}(\rho)\leq QC_M^{\searrow 0} (\rho)+2\lfloor\log
k\rfloor+2, 
\qquad \rho \in \mathcal{T}^+_1(\hr_{\Fock} ).
\label{eqRelation}
\end{equation}
\end{lemma}
{\bf Proof. } Suppose that $QC_M^{\searrow 0}(\rho)=l$, so there is
a density matrix $\sigma \in \mathcal{T}^+_1(\hr_{\Fock} )$ with 
$\ell(\sigma)=l$, such that $\|M(k,\sigma)-\rho\|_{\rm Tr}\leq \frac 1 k$ for every
$k\in\N$. Then $\sigma':= \mathcal{C}(k,\sigma)$, where $\mathcal{C}$ 
is given in Definition \ref{DefEncoding}, is an input for $M$ such
that $\|M(\sigma')-\rho \|_{\rm Tr}\leq \frac 1 k$. 
Thus  $QC_M^{1/k}(\rho)\leq \ell(\sigma') \leq 2\lfloor\log k\rfloor
+2+ \ell(\sigma)= 2\lfloor\log k\rfloor +2+ QC_M^{\searrow 0} (\rho)$, 
where the second inequality is by (\ref{encoding_length}).
\qed

The term $2\lfloor\log k\rfloor+2$ in (\ref{eqRelation}) depends on
our encoding $\mathcal{C}$ given in Definition~\ref{DefEncoding},
but if $M$ is assumed to be universal
(which will be discussed below), then (\ref{eqRelation}) will hold
for {\em every} encoding, if we replace the term $2\lfloor\log
k\rfloor+2$ by $K(k)+c_M$, where 
$K(k)\leq 2\lfloor\log k\rfloor+\mathcal{O}(1)$
denotes the classical (self-delimiting)
algorithmic complexity of the integer $k$, and $c_M$ is some
constant depending only on $M$. For more details we refer the reader
to~\cite{Vitanyibook}.

In \cite{BernsteinVazirani}, it is proved that there is a universal
{\sl QTM} ({\sl UQTM}) $\mathfrak{U}$ that can simulate with
arbitrary accuracy every other machine $M$ in the sense that for
every such $M$ there is a classical bit string $\bar M\in\{0,1\}^*$ such that
\begin{eqnarray}
\|
\mathfrak{U}(\bar M,\sigma,k,t)-M^t_{\rm\bf O}(\sigma)
\|_{\rm Tr}\leq \frac 1 k\qquad \mbox{for every }
\sigma\in\mathcal{T}_1^+(\hr_{\Fock}),
   \label{eqSimulation}
\end{eqnarray}
where $k,t\in\N$. As it is implicit in this
definition of universality, we will demand that $\mathfrak U$ is
able to perfectly simulate every classical computation, and that it
can apply a given unitary transformation within any desired accuracy
(it is shown in \cite{BernsteinVazirani} that such machines exist).

We choose an arbitrary {\sl UQTM} $\mathfrak U$ which is constructed such
that it decodes our encoding $\mathcal{C}(k,\sigma)$ given in 
Definition~\ref{DefEncoding}
into $k$ and $\sigma$ at the beginning of the computation.
Like in the classical case, we fix $\mathfrak U$ for the rest of the paper
and simplify notation by
\[
   QC^{\searrow 0}(\rho):=QC_{\mathfrak U}^{\searrow 0}(\rho),\qquad
   QC^\delta(\rho):=QC_{\mathfrak U}^\delta(\rho)\,\,.
\]

\section{Proof of the Main Theorem}
As already mentioned at the beginning of Section~\ref{SecErgodicQuantumSources},
without loss of generality, we give the proofs for
the case that $\mathcal{A}$ is the algebra of the observables of a qubit,
i.e. the
complex $2\times 2$-matrices.

\subsection{Lower Bound}
For classical {\sl TM}s, there are no more than $2^{c+1}-1$
different programs of length $\ell\leq c$.
This can be used as a ''counting argument'' for proving
the lower bound of Brudno's Theorem in the classical case (\cite{Keller}).
We are now going to prove a similar statement for {\sl QTM}s.

Our first step is to elaborate on an argument due
to~\cite{Berthiaume} which states that there cannot be more than $2^{\ell+1}-1$
mutually orthogonal one-dimensional projectors $p$ with quantum complexity
$QC^{\searrow 0}(p)\leq\ell$.
The argument is based on Holevo's $\chi$-quantity
associated to any ensemble
$\mathbb{E}_\rho:=\left\{\lambda_i,\rho_i\right\}_i$ consisting of weights
$0\leq\lambda_i\leq 1$, $\sum_i\lambda_i=1$, and of density matrices
$\rho_i$ acting on a Hilbert space $\mathcal{H}$.
Setting $\rho:=\sum_i\lambda_i\rho_i$, the $\chi$-quantity is defined as
follows
\begin{eqnarray}
\label{eqHolevoLogDim1}
\chi(\mathbb{E}_\rho)&:=&S\left(\rho\right)\,-\,\sum_i\lambda_i\,S(\rho_i)\\
  &=&\sum_i\lambda_i\,S(\rho_i,\rho)\ ,
   \label{eqHolevoLogDim2}
\end{eqnarray}
where, in the second line, the relative entropy appears
\begin{equation}
   \label{relent}
    S(\rho_1\,,\,\rho_2)
   :=\left\{
      \begin{array}{cl}
          {\rm
    Tr}\Bigl(\rho_1\Bigl(\log\rho_1\,-\,\log\rho_2\Bigr)\Bigr) 
& \mbox{if }
          {\rm supp }\enspace\rho_1\leq {\rm supp }\enspace\rho_2\,\,,\\
          \infty & \mbox{otherwise}.
      \end{array}
   \right.
\end{equation}
If $\hbox{dim}(\mathcal{H})$ is finite,~(\ref{eqHolevoLogDim1}) is
bounded by the maximal von Neumann entropy:
\begin{equation}
\label{eqHolevoLogDim3}
\chi(\mathbb{E}_\rho)\leq S(\rho) \leq \log{\hbox{dim}(\hr)}.
\end{equation}

In the following, $\hr'$ denotes an arbitrary (possibly infinite-dimensional)
Hilbert space, while the rest of the notation is adopted from
Subsection~\ref{BasicDefQTMSec}.

\begin{lemma}[Quantum Counting Argument]\label{CountingArgument}
Let $0<\delta<1/e$, $c\in\N$ such that $c\geq \frac 2 \delta\left(2+\log\frac 1
\delta\right)$, $P$ an orthogonal projector onto a linear subspace of an arbitrary Hilbert space
$\hr'$, and
$\mathcal{E}:\mathcal{T}^+_1(\hr_{\Fock})\to\mathcal{T}^+_1(\hr')$ a
quantum operation.
Let $N^\delta_c$ be a subset of
one-dimensional mutually orthogonal projections from the set
\begin{eqnarray*}
A_c^\delta(\mathcal{E},P)
:=\{p \leq P\ |\ p \textrm{ 1-dim. proj.}, \exists \sigma
\in \mathcal{T}^+_1(\hr_{\leq c}):
\|\mathcal{E}(\sigma)- p\|_{\rm Tr}
\leq \delta \},
\end{eqnarray*}
that is, the set of all pure quantum states which are reproduced 
within $\delta$ by the operation $\mathcal{E}$ on some input of length 
$\leq c$.
Then it holds that
\begin{eqnarray*}
\log | N^\delta_c | < c+1+\frac{2+\delta}{1-2\delta}\delta c\,\,.
\end{eqnarray*}
\end{lemma}
{\bf Proof. } Let $p_j\in
A_c^\delta(\mathcal{E},P)$, $j=1,\ldots,N$, be a set of mutually orthogonal
projectors 
 and $p_{N+1}:={\bf
1}_{\hr'}-\sum_{i=1}^N p_i$. 
By the definition of $A_c^\delta(\mathcal{E},P)$, for every $1\leq i \leq N$,
there are density matrices $\sigma_i\in\mathcal{T}_1^+(\hr_{\leq c})$
with
\begin{eqnarray}\label{delta_distance}
   \|\mathcal{E}(\sigma_i)-p_i\|_{\rm Tr}\leq \delta\,\,.
\end{eqnarray}
Consider the equidistributed ensemble
$\mathbb{E}_\sigma:=\Bigl\{\frac{1}{N},\sigma_i\Bigr\}$,
where $\displaystyle\sigma:=\frac{1}{N}\sum_{i=1}^N\sigma_i$ also acts
on $\hr_{\leq c}$. Using that $\dim \hr_{\leq c} = 2^{c+1}-1$, 
inequality (\ref{eqHolevoLogDim3}) yields
\begin{eqnarray}\label{dim_estim}
\chi(\mathbb{E}_\sigma)< c+1.
\end{eqnarray}
We define  a quantum operation ${\mathcal R}$ on 
$\mathcal{T}^+_1(\hr')$ by 
${\mathcal R} (a)
:= \sum_{i=1}^{N+1} p_i a p_i$.
Applying twice the monotonicity of the relative entropy under quantum 
operations, we obtain
\begin{equation}
\frac{1}{N} \sum_{i=1}^{N}\,
S\left({\mathcal R} \circ \mathcal{E}(\sigma_i)
, {\mathcal {R}} \circ \mathcal{E}(\sigma)
\right)\leq
\frac{1}{N} \sum_{i=1}^{N}\,
S\left(\mathcal{E}(\sigma_i), \mathcal{E}(\sigma)\right)
\leq\chi(\mathbb{E}_\sigma)\,\,.
\label{chi_estimate}
\end{equation}
Moreover, for every $i\in\{1, \dots, N\}$, the density operator 
$\mathcal{R}\circ \mathcal{E}(\sigma_i)$ is close to the 
corresponding one-dimensional projector 
$\mathcal{R}(p_i)=p_i$. Indeed, by the contractivity 
of the trace distance under quantum operations (compare Thm. 9.2 in
\cite{NielsenChuang}) and by assumption (\ref{delta_distance}), it holds
\begin{eqnarray*}
   \|\mathcal{R}\circ \mathcal{E}(\sigma_i)
   - p_i\|_{\rm Tr}\leq \|\mathcal{E}(\sigma_i)-
   p_i\|_{\rm Tr}
\leq \delta\,\,.
\end{eqnarray*}
Let $\Delta:=\frac 1 N\sum_{i=1}^N p_i$. 
The trace-distance is convex (\cite{NielsenChuang}, (9.51)), thus
$$
\|\mathcal{R}\circ \mathcal{E}(\sigma)-\Delta\|_{\rm Tr}\leq
\frac 1 {N} \sum_{i=1}^{N} \|\mathcal{R}\circ \mathcal{E}(\sigma_i)
-p_i\|_{\rm Tr}\leq \delta\,\,,
$$
whence, since $\delta<\frac 1 e$, Fannes' inequality 
(compare Thm. 11.6 in \cite{NielsenChuang}) gives
\begin{eqnarray*}
   \label{eqFannes1}
   &&S(\mathcal{R}\circ \mathcal{E}(\sigma_i))=\left\vert
   S(\mathcal{R}\circ \mathcal{E}(\sigma_i))-S(p_i)\right\vert
   \leq \delta\log(N+1)+\eta(\delta)\\
   \mbox{and }&&\left\vert S(\mathcal{R}\circ \mathcal{E}(\sigma)
)-S(\Delta)\right\vert
   \leq\delta\log(N+1)+\eta(\delta)\,\,,
   \label{eqFannes2}
\end{eqnarray*}
where $\eta(\delta):=-\delta\log\delta$.
Combining the two estimates above with (\ref{dim_estim}) and 
(\ref{chi_estimate}), we obtain
\begin{eqnarray}
   c+1&>& \chi(\mathbb{E}_\sigma)\geq
   S(\mathcal{R}\circ \mathcal{E}(\sigma))-\frac 1 {N} 
\sum_{i=1}^{N} S(\mathcal{R}\circ \mathcal{E}(\sigma_i)) 
\nonumber \\
\nonumber &\geq& S(\Delta)-\delta\log(N+1)-\eta(\delta)
-\frac 1 N \sum_{i=1}^N\left(\delta\log(N+1)+\eta(\delta)\right)\nonumber\\
&=&\log N-2\delta \log(N+1)-2\eta(\delta) \nonumber\\
&\geq&(1-2\delta)\log N-2\delta-2\eta(\delta).
\label{constr}
\end{eqnarray}
Assume now that $\log N\geq
c+1+\frac{2+\delta}{1-2\delta}\delta c$. Then it follows (\ref{constr}) that
$c<\frac 2 \delta\left(2+\log\frac 1 \delta\right)$.
So if $c$ is larger than this expression, the maximum number
$| N^\delta_c |$ of mutually orthogonal projectors in $A^\delta_c(\mathcal{E}, P)$
must be bounded by $\displaystyle \log | N^\delta_c | <
c+1+\frac{2+\delta}{1-2\delta}\delta c$.\qed

The second step uses the previous lemma together with the
following theorem \cite[Prop. 2.1]{QSMPaper}. It is closely related to the quantum
Shannon-McMillan Theorem and concerns the minimal
dimension of the $\Psi-$typical subspaces.

\begin{theorem}\label{QAEP}
Let $(\mathcal{A}^\infty,\Psi)$ be an ergodic quantum source with
entropy rate $s$.
Then, for every $0<\varepsilon<1$,
\begin{equation}
   \lim_{n\to\infty}\frac 1 n \beta_{\eps,n}(\Psi)=s,
   \label{eqBoltzmann}
\end{equation}
where $
   \beta_{\eps,n}(\Psi):=\min\left\{
      \log {\rm Tr}_n(q)\enspace|\enspace q\in\mathcal{A}^{(n)}
\mbox{ projector },
      \Psi^{(n)}(q)\geq 1-\eps
   \right\}$.
\end{theorem}
Notice that the limit (\ref{eqBoltzmann}) is valid for all
$0<\eps<1$. By means of this property, we will first prove the lower
bound for the finite-accuracy complexity $QC^\delta$, and then use
Lemma~\ref{LemRelation} to extend it to $QC^{\searrow 0}$.
\begin{corollary}[Lower Bound for $\frac 1 n QC^\delta$]
\label{Cor1}
\lineclear
Let $(\mathcal{A}^\infty,\Psi)$ be an
ergodic quantum source with entropy rate $s$. Moreover, let
$0<\delta<1/e$, and let
$\left(p_n\right)_{n\in\N}$ be a sequence of $\Psi$-typical projectors.
Then, there is another sequence of $\Psi$-typical projectors
$q_n(\delta)\leq p_n$, such that for $n$ large enough
\[
\frac{1}{ n} QC^{\delta}(q)>s-\delta(2+\delta)s\]
is true for every one-dimensional projector $q\leq q_n(\delta)$.
\end{corollary}
{\bf Proof. } The case $s=0$ is trivial, so let $s>0$.
Fix $n \in \N$ and some $0<\delta<1/e$, and consider the set
\[
   \tilde{A}_n(\delta):=\left\{
      p\leq p_n\ |\ p \textrm{ one-dim. proj., }
      QC^\delta(p)\leq ns(1-\delta(2+\delta))\right\}.
\]

From the definition of $QC^\delta(p)$, to
all $p$'s there exist associated density matrices $\sigma_p$ with
$\ell(\sigma_p)\leq ns(1-\delta(2+\delta))$
such that
$\|\mathcal{M}(\sigma_p)-p\|_{\rm Tr}\leq \delta$, where $\mathcal{M}$ denotes
the quantum operation $\mathcal{M}:\mathcal{T}_1^+(\hr_{\Fock})\to
\mathcal{T}_1^+(\hr_\Fock)$ of the corresponding {\sl UQTM} $\mathfrak{U}$,
as explained in Lemma~\ref{LemmaQTMsAreOperations}.
Using the notation of Lemma~\ref{CountingArgument}, it thus follows that
$$
   \tilde A_n(\delta)
   \subset A_{\lceil ns(1-\delta(2+\delta))\rceil}^\delta(\mathcal{M},
   p_n)\,\,.
$$
Let $p_n(\delta)\leq p_n$ be a sum of
a maximal number of mutually orthogonal projectors from
 $A^\delta_{\lceil ns(1-\delta(2+\delta))\rceil}(\mathcal{M},p_n)$.
If $n$ was chosen large enough such that $\label{N_tilde}
ns(1-\delta(2+\delta))\geq \frac 1 \delta\left(4+2\log\frac 1
\delta\right)$ 
is satisfied, Lemma~\ref{CountingArgument} implies that
\begin{eqnarray}
   \log {\rm Tr}\,p_n(\delta)< \lceil ns(1-\delta(2+\delta))\rceil +1+\frac{2+\delta}{1-2\delta}
   \delta \lceil n s (1-\delta(2+\delta))\rceil,
   \label{eqCountArgUsed}
\end{eqnarray}
and there are no one-dimensional projectors $p \leq p_n(\delta)^\perp:= p_n - p_n(\delta)$ such that 
$p \in A_{\lceil ns(1-\delta(2+\delta))\rceil }^\delta(\mathcal{M}, p_n)$,
Namely, 
one-dimensional projectors $p \leq p_n(\delta)^{\perp}$ must satisfy
$\frac{1}{n}QC^{\delta}(p)> s- \delta(2+\delta)s$.
Since inequality (\ref{eqCountArgUsed}) is
valid for every $n\in\N$ large enough, we conclude
\begin{eqnarray}
   \limsup_{n\to\infty}\frac 1 n \log {\rm Tr}_n p_n(\delta)
   \leq s-2\delta^3 s -\frac{5\delta^4 s}{1-2\delta}<s.
   \label{eqSmallerS}
\end{eqnarray}
Using Theorem~\ref{QAEP}, we obtain that
$\lim_{n\to\infty}\Psi^{(n)}(p_n(\delta))=0$.
Finally, set $q_n(\delta):=p_n(\delta)^\perp$.
The claim follows. \qed

\begin{corollary}[Lower Bound for $\frac 1 n QC^{\searrow 0}$]
\label{Cor2}
\lineclear
Let $(\mathcal{A}^\infty,\Psi)$ be an
ergodic quantum source with entropy rate $s$.
Let $\left(p_n\right)_{n\in\N}$ with $p_n\in\mathcal{A}^{(n)}$ be an
arbitrary sequence of $\Psi$-typical projectors. 
Then, for every $0<\delta <1/e$,
there is a sequence of $\Psi$-typical projectors $q_n(\delta)\leq p_n$ such
that for $n$ large enough
\[
\frac 1 n QC^{\searrow 0}(q)>s-\delta
\]
is satisfied for every one-dimensional projector
$q\leq q_n(\delta)$.
\end{corollary}
{\bf Proof. } According to Corollary~\ref{Cor1}, for
every $k \in \N$, there exists a sequence of $\Psi$-typical 
projectors $p_n(\frac 1 k)\leq p_n$
with $\frac 1 n QC^\frac 1 k (q)>s-\frac 1 k (2+\frac 1 k)s$ for
every one-dimensional projector $q\leq p_n(\frac 1 k )$ if $n$ is large enough.
We have
\begin{eqnarray*}
   \frac 1 n QC^{\searrow 0}(q)&\geq&\frac 1 n QC^{1/k}(q)-
   \frac {2+2\lfloor\log k\rfloor}{n}\\
   &>& s-\frac 1 k \left(2+\frac 1 k\right)s-\frac{2(2+\log k)} n,
\end{eqnarray*}
where the first estimate is by Lemma~\ref{LemRelation}, and the second
one is true for one-dimensional projectors $q \leq p_n(\frac 1 k )$
and $n\in\N$ large enough. Fix a large $k$ satisfying 
$\frac 1 k (2 + \frac 1 k )s \leq \frac \delta 2 $. 
The result follows by setting $q_n(\delta)=p_n(\frac 1 k)$.
\qed

\subsection{Upper Bound}
In the previous section, we have shown that with high probability
and for large $m$, the finite-accuracy complexity rate
$\frac 1 m QC^{\delta}$ is bounded from
below by $s(1-\delta(2+\delta))$, and the approximation-scheme
quantum complexity rate $\frac 1 m QC^{\searrow 0}$ by $s-\delta$.
We are now going to establish the upper bounds.
\begin{proposition}[Upper Bound]
\label{PropUpperBound}
\lineclear
Let $(\mathcal{A}^\infty,\Psi)$ be an ergodic quantum source with entropy rate
$s$.
Then, for every $0<\delta<1/e$, there is a sequence of 
$\Psi$-typical projectors $q_m(\delta)\in\mathcal{A}^{(m)}$ such that
for every one-dimensional
projector $q\leq q_m(\delta)$ and $m$ large enough
\begin{eqnarray}
   \frac 1 m QC^{\searrow 0}(q)&<&s+\delta\qquad\mbox{and}\label{UpperZero}\\
   \frac 1 m QC^\delta(q)&<&s+\delta\,\,.\label{UpperDelta}
\end{eqnarray}
\end{proposition}
We prove the above proposition by explicitly providing a quantum
algorithm (with program length increasing like $m(s+\delta)$)
that computes $q$ within arbitrary accuracy. This will be done by
means of quantum universal typical subspaces constructed by Kaltchenko and Yang in \cite{KaltchenkoYang}. 
\begin{theorem}[Universal Typical Subspaces]\label{Kaltc}
\lineclear
Let $s > 0$ and $\varepsilon>0$. There exists a sequence of
projectors $Q^{(n)}_{s,\varepsilon} \in \mathcal{A}^{(n)} $, $n\in\N$, 
such that for $n$ large enough
\begin{eqnarray}
\displaystyle
{\rm Tr}\Bigl(Q^{(n)}_{s,\varepsilon}\Bigr)\leq
2^{n(s+\varepsilon)}\label{eqTraceIsSmall}
\end{eqnarray}
and  for every ergodic quantum state $\Psi \in \mathcal{S}(\mathcal{A}^\infty)$
with entropy rate $s(\Psi) \leq s$ it holds that
\begin{eqnarray}
\lim_{n\to\infty}\Psi^{(n)}(Q^{(n)}_{s,\varepsilon})=1\,\,.
\end{eqnarray}
\end{theorem}
We call the orthogonal projectors $Q_{s,\varepsilon}^{(n)}$ in the
above theorem universal typical projectors at level $s$.
Suited for designing an appropriate quantum algorithm,
we slightly modify the proof given by Kaltchenko and
Yang in \cite{KaltchenkoYang}. 

{\bf Proof. } Let $l \in \mathbb{N}$ and $R > 0$. We consider an Abelian
quasi-local subalgebra $\mathcal{C}_{l}^\infty \subseteq
\mathcal{A}^\infty $ constructed from a maximal Abelian $l-$block
subalgebra $\mathcal{C}_{l}\subseteq \mathcal{A}^{(l)} $. The results in
\cite{Ziv,Kieffer} imply that there exists a universal sequence 
of projectors $p_{l,R}^{(n)} \in \mathcal{C}^{(n)}_l 
\subseteq \mathcal{A}^{(ln)}$ with $\frac{1}{n} \log \textrm{Tr }
p_{l,R}^{(n)}\leq R$ such that $\lim_{n \to \infty} \pi^{(n)}(p_{l,
R}^{(n)})=1$ 
for any ergodic state $\pi$ on the Abelian algebra $\mathcal{C}_l^{\infty}$
with entropy rate $s(\pi) < R$. Notice that ergodicity and entropy
rate of $\pi$ are defined with respect to
the shift on $\mathcal{C}_l^\infty$, which corresponds to
the $l$-shift on $\mathcal{A}^{\infty}$.

The first step in \cite{KaltchenkoYang} is to
apply unitary operators of the form $U ^{\otimes n}$,
$U\in \mathcal{A}^{(l)}$
unitary, to the $p_{l,R}^{(n)} $ and to introduce the projectors
\begin{eqnarray}
w_{l,R}^{(ln)}:=\bigvee_{U\in\mathcal{A}^{(l)}\mbox{ unitary}}
   U^{\otimes n} p_{l,R}^{(n)} U^{*\otimes n} \in \mathcal{A}^{(ln)}.
   \label{eqJoin1}
\end{eqnarray}
Let $p_{l,R}^{(n)}=\sum_{i\in I}|i_{l,R}^{(n)}\rangle\langle i_{l,R}^{(n)}|$ be
a spectral decomposition of $p_{l,R}^{(n)}$ (with $I\subset\N$ some
index set), and let $\mathbf{P}(V)$ denote the orthogonal projector
onto a given subspace $V$.
Then, $w_{l,R}^{(ln)}$ can also be written as
\[
   w_{l,R}^{(ln)}=\mathbf{P}\left(
      {\rm span}\{
         U^{\otimes n}|i_{l,R}^{(n)}\rangle:
         i\in I, U\in\mathcal{A}^{(l)}\mbox{ unitary}
      \}
   \right).
\]
It will be more convenient for the construction of our algorithm 
in \ref{subsubconstr} to consider the projector
\begin{equation}
   W_{l,R}^{(ln)}:=\mathbf{P}\left(
      {\rm span}\{
         A^{\otimes n}|i_{l,R}^{(n)}\rangle:
         i\in I, A\in\mathcal{A}^{(l)}
      \}
   \right).
   \label{eqJoin2}
\end{equation}
It holds that $w_{l,R}^{(ln)}\leq W_{l,R}^{(ln)}$.
For integers $m = nl + k$ with $n \in \mathbb{N}$ and $k \in \{0,\dots, l-1\}$
we introduce the projectors in $\mathcal{A}^{(m)}$
\begin{eqnarray}
   w_{l,R}^{(m)}:=w_{l,R}^{(ln)} \otimes \mathbf{1}^{\otimes k},\qquad
   W_{l,R}^{(m)}:=W_{l,R}^{(ln)} \otimes \mathbf{1}^{\otimes k}.
\end{eqnarray}
We now use an argument of \cite{JHHH} to estimate the trace of
$W_{l,R}^{(m)} \in \mathcal{A}^{(m)}$. The dimension of the symmetric subspace
$\textrm{SYM}^n(\mathcal{A}^{(l)}):={\rm span}\{A^{\otimes
n}:A\in\mathcal{A}^{(l)}\}$ is upper bounded by $(n+1)^{\dim
\mathcal{A}^{(l)}}$, thus
\begin{eqnarray}
   \hbox{Tr } W_{l,R}^{(m)} = \hbox{Tr } W_{l,R}^{(ln)}
   \cdot  \hbox{Tr }\mathbf{1}^{\otimes k}  &\leq& (n+1)^{2^{2l}}
   \hbox{Tr }p_{l,R}^{(n)} \cdot 2^{l} \nonumber\\&\leq&  (n+1)^{2^{2l}}
   \cdot 2^{Rn} \cdot 2^{l} \label{dim_estimate}.
\end{eqnarray}
Now we consider a stationary ergodic state $\Psi$  on the
quasi-local algebra $\mathcal{A}^\infty$ with entropy rate $s(\Psi)
\leq s$. Let $\eps, \delta > 0$. If  $l$ is chosen large enough then the
projectors $w_{l,R}^{(m)}$, where
$R:=l(s+ \frac{\eps}{2})$, are $\delta-$typical for $\Psi$, i.e. 
$\Psi^{(m)}(w_{l,R}^{(m)})\geq 1- \delta$, for $m \in \N$ sufficiently
large. This can be seen as
follows. Due to the result in \cite[Thm. 3.1]{QSMPaper} the ergodic state
$\Psi$ convexly decomposes into $k(l)\leq l$ states
\begin{eqnarray}\label{erg_decomp}
\Psi = \frac{1}{k(l)}\sum_{i=1}^{{k(l)}} \Psi_{i,l},
\end{eqnarray}
each $\Psi_{i,l}$ being ergodic with respect to the $l-$shift on
$\mathcal{A}^\infty$ and having an entropy rate (with respect to the
$l-$shift) equal to $s(\Psi)\cdot l$. 
We define for $\Delta >0$ the set of integers
\begin{eqnarray}
A_{l, \Delta}:= \{i \in \{1, \dots, k(l)\}: \ S(\Psi^{(l)}_{i,l})
\geq l(s(\Psi)+ \Delta) \}.
\end{eqnarray}
Then, according to a density lemma proven in \cite[Lemma 3.1]{QSMPaper} it holds
\begin{eqnarray}
\lim_{l \to \infty} \frac{|A_{l,\Delta}|}{k(l)}=0.
\end{eqnarray}
Let $\mathcal{C}_{i,l}$ be the maximal Abelian subalgebra of
$\mathcal{A}^{(l)}$ generated by the one-dimensional
eigenprojectors of $\Psi_{i,l}^{(l)}\in
\mathcal{S}(\mathcal{A}^{(l)} )$.
The restriction of a component $\Psi_{i,l}$ to the Abelian
quasi-local algebra $\mathcal{C}_{i,l}^\infty $ is again an ergodic
state. It holds in general
\begin{eqnarray}
l \cdot s(\Psi) = s(\Psi_{i,l})\leq s(\Psi_{i,l} \upharpoonright
\mathcal{C}_{i,l}^\infty)  \leq
S(\Psi_{i,l}^{(l)}\upharpoonright \mathcal{C}_{i,l})
=S(\Psi^{(l)}_{i,l}).
\end{eqnarray}
For $i \in A_{l,\Delta}^c$, where we set $\Delta:=\frac{R}{l}-s(\Psi)$,
we additionally have
the upper bound $S(\Psi^{(l)}_{i,l}) < R$.
Let $U_i \in \mathcal{A}^{(l)}$ be a unitary operator such that
$U_i^{\otimes n} p_{l,R}^{(n)}U_i^{*\otimes n}
\in \mathcal{C}_{i,l}^{(n)}$.
For every $i \in A_{l,\Delta}^c$, it holds that
 \begin{eqnarray}\label{erg_comp}
\Psi_{i,l}^{(ln)}(w_{l,R}^{(ln)}) \geq
\Psi_{i,l}^{(ln)}(U_i^{\otimes n} p_{l,R}^{(n)}U_i^{*\otimes n})
\longrightarrow 1\qquad
\textrm{as } n \to \infty.
\end{eqnarray}
We fix an $l \in \N$ large enough to fulfill
$\frac{|A_{l,\Delta}^c|}{k(l)}\geq 1-\frac{\delta}{2}$ and use the
ergodic decomposition  (\ref{erg_decomp})
 to obtain the lower bound
\begin{eqnarray}
\Psi^{(ln)}(w_{l,R}^{(ln)})\geq \frac{1}{k(l)}
\sum_{i \in A_{l,\Delta}^c}\Psi_{l,i}^{(nl)}(w_{l,R}^{(ln)})
\geq \left(1- \frac{\delta}{2}\right)\min_{i \in A_{l,\Delta}^c}
\Psi_{i,l}^{(nl)}(w_{l,R}^{(ln)}).
\end{eqnarray}
\,From (\ref{erg_comp}) we conclude that for $n$ large enough
\begin{eqnarray}
   \Psi^{(ln)}(W_{l,R}^{(ln)})\geq \Psi^{(ln)}(w_{l,R}^{(ln)}) \geq 1-\delta.
\end{eqnarray}
We proceed by following the lines of \cite{KaltchenkoYang} by
introducing  the sequence $l_m$, $m \in \N$, where each $l_m$ 
is a power of $2$ fulfilling the inequality
\begin{eqnarray}\label{l_m}
l_m 2^{3\cdot l_m} \leq m < 2 l_m 2^{3\cdot 2 l_m}.
\end{eqnarray}
Let the integer sequence $n_m$ and the real-valued sequence $R_m$ be defined
 by $n_m:=\lfloor \frac m {l_m}\rfloor$ and 
$R_m:=l_m \cdot \left(s+ \frac \eps 2\right)$. Then we set
\begin{equation}
   Q_{s,\eps}^{(m)}:=\left\{
      \begin{array}{ll}
          W_{l_m,R_m}^{(l_m n_m)} & \mbox{if }m=l_m 2^{3\cdot l_m}\,\,,\\
          W_{l_m,R_m}^{(l_m n_m)}\otimes \idn^{\otimes(m-l_m n_m)}
          & \mbox{otherwise}\,\,.
      \end{array}
   \right.
   \label{eqTypicalProjector}
\end{equation}
Observe that
\begin{eqnarray}
\frac{1}{m}\log \hbox{Tr }Q_{s,\eps}^{(m)} & \leq& \frac{1}{n_ml_m}
\log \hbox{Tr }Q_{s,\eps}^{(m)}
\nonumber \\ 
&\leq& \frac{4^{l_m}}{l_m}\frac{\log(n_m+1)}{n_m}
+\frac{R_m}{l_m}+\frac{1}{n_m}\\ 
&\leq & \frac{4^{l_m}}{l_m}\frac{6l_m +2}{2^{3l_m}-1}
+ s+\frac{\eps}{2} +\frac{1}{2^{3l_m}-1},
\end{eqnarray}
where the second inequality is by estimate (\ref{dim_estimate}) 
and the last one by the bounds on $n_m$
\begin{eqnarray*}
2^{3l_m}-1\leq \frac{m}{l_m}-1 \leq n_m \leq \frac{m}{l_m}\leq 2^{6l_m +1}.
\end{eqnarray*}
Thus, for large $m$, it holds
\begin{eqnarray}
   \frac{1}{m} \log \textrm{Tr }Q_{s,\eps}^{(m)} \leq s +\eps.
   \label{eqLogTrQm}
\end{eqnarray}
By the special choice (\ref{l_m}) of $l_m$ it is ensured that the sequence of
projectors $Q_{s,\eps}^{(m)} \in \mathcal{A}^{(m)}$ is indeed typical for
any quantum state $\Psi$ with entropy rate $s(\Psi) \leq s$, 
compare \cite{KaltchenkoYang}. This means that
$\{Q_{s,\eps}^{(m)}\}_{m \in \mathbf{N}}$ is a sequence of universal
typical projectors at level $s$. \qed
\subsubsection{Construction of the Decompression Algorithm}
\label{subsubconstr}
We proceed by applying the latter result to universal typical subspaces
for our proof of the upper bound. Let
$0<\eps<\delta/2$ be an arbitrary real number such that $r:=s+\eps$ is
rational, and let $q_m:=Q_{s,\eps}^{(m)}$ be the universal
projector sequence of Theorem~\ref{Kaltc}. Recall that
the projector sequence $q_m$ is {\em independent} of the choice of
the ergodic state $\Psi$, as long as $s(\Psi)\leq s$.

Because of (\ref{eqTraceIsSmall}), for $m$ large enough,
there exists some unitary transformation $U^*$ that transforms the projector
$q_m$ into a projector belonging to $\mathcal{T}_1^+(\hr_{\lceil mr\rceil})$,
thus transforming every one-dimensional projector
$q\leq q_m$ into a qubit string $\tilde q:=U^* q U$ of length 
$\ell(\tilde q)=\lceil mr\rceil$.

As shown in \cite{BernsteinVazirani}, a {\sl UQTM} can implement
every classical algorithm, and it can apply every unitary
transformation $U$ (when given an algorithm for the computation of $U$) on
its tapes within any desired accuracy.
We can thus feed $\tilde q$ (plus some classical instructions
including a subprogram for the computation of $U$) as
input into the {\sl UQTM} $\mathfrak U$. This {\sl UQTM} starts by
computing a classical description of the transformation $U$, and
subsequently applies $U$ to $\tilde q$, recovering the original
projector $q=U\tilde q U^*$ on the output tape.

Since $U=U(q_m)$ depends on $\Psi$ only through
its entropy rate $s(\Psi)$,
the subprogram that computes $U$ does not have to be
supplied with additional information on $\Psi$ and will thus
have fixed length.

We give a precise definition of a quantum decompression algorithm 
$\mathfrak{A}$, which is, formally, a mapping ($r$ is rational)
\begin{eqnarray*}
   \mathfrak{A}:\N\times\N
   \times{\mathbb Q}\times\hr_{\Fock}\to\hr_{\Fock}\,\,,\\
   (k,m,r,\tilde q)\mapsto      q=\mathfrak{A}(k,m,r,\tilde q)\,\,.
\end{eqnarray*}
We require that $\mathfrak{A}$ is a  ''short algorithm'' in the sense 
of ''short in description'', {\em not}
short (fast) in running time or resource consumption. Indeed, the algorithm
$\mathfrak{A}$ is very slow and memory consuming, but this does
not matter, since Kolmogorov complexity only cares about the
description length of the program.

The instructions defining the quantum algorithm $\mathfrak{A}$ are:
\begin{itemize}
\item[1.] Read the value of $m$, and find a solution $l\in\N$ for the 
inequality
\[
   l \cdot 2^{3 l}\leq m < 2 \cdot l\cdot 2^{3\cdot 2 l}
\]
such that $l$ is a power of two. (There is only one such $l$.)
\item[2.]Compute $n:=\lfloor\frac m l\rfloor$.
\item[3.] Read the value of $r$. Compute $R:=l\cdot r$.
\item[4.] Compute a list of codewords $\Omega_{l,R}^{(n)}$,
belonging to a classical universal block
code sequence of rate $R$. (For the construction of an appropriate algorithm,
see \cite[Thm. 2 and 1]{Kieffer}.) Since
\[
   \Omega_{l,R}^{(n)}\subset\left(\{0,1\}^l\right)^n\,\,,
\]
$\Omega_{l,R}^{(n)}=\{\omega_1,\omega_2,\ldots,\omega_M\}$ 
can be stored as a list
of binary strings. Every string has length $\ell(\omega_i)=nl$. 
(Note that the exact value
of the cardinality $M\approx 2^{nR}$ depends on the choice of 
$\Omega_{l,R}^{(n)}$.)
\end{itemize}
During the following steps, the quantum algorithm $\mathfrak{A}$
will have to deal with
\begin{itemize}
\item rational numbers,
\item square roots of rational numbers,
\item binary-digit-approximations (up to some specified accuracy) of
real numbers,
\item (large) vectors and matrices containing such numbers.
\end{itemize}
A classical {\sl TM} can of course deal with all such objects
(and so can {\sl QTM}):
For example, rational numbers can be stored
as a list of two integers (containing numerator and denominator),
square roots can be stored as such a list and an additional bit
denoting the square root, and binary-digit-approximations can be
stored as binary strings. Vectors and matrices are arrays containing
those objects. They are always assumed to be given in the
computational basis. Operations on those objects, like addition or
multiplication, are easily implemented.

The quantum algorithm $\mathfrak{A}$ continues as follows:
\begin{itemize}
\item[5.] Compute a basis $\left\{A_{\{i_1,\ldots,i_n\}}\right\}$
of the symmetric subspace
\[
   \textrm{SYM}^n(\mathcal{A}^{(l)}):={\rm span}\{A^{\otimes n}:A
   \in\mathcal{A}^{(l)}\}\,\,.
\]
This can be done as follows: For every $n$-tuple $\{i_1,\ldots,i_n\}$,
where $i_k\in\{1,\ldots,2^{2l}\}$,
there is one basis element $A_{\{i_1,\ldots,i_n\}}\in\mathcal{A}^{(ln)}$,
given by the formula
\begin{equation}
   A_{\{i_1,\ldots,i_n\}}=\sum_{\sigma}
   e^{(l,n)}_{\sigma(i_1,\ldots,i_n)}\,\,,
   \label{eqWillBeRational}
\end{equation}
where the summation runs over all $n$-permutations $\sigma$, and
\[
   e_{i_1,\ldots,i_n}^{(l,n)}:=e_{i_1}^{(l)}\otimes
   e_{i_2}^{(l)}\otimes\ldots
   \otimes e_{i_n}^{(l)}\,\,,
\]
with $\left\{e_k^{(l)}\right\}_{k=1}^{2^{2l}}$ a system of matrix
units\footnote{In the computational basis, all entries are zero,
except for one entry which is one.} in $\mathcal{A}^{(l)}$.

There is a number of $d={{n+2^{2l}-1}\choose {2^{2l}-1}}
=\dim(\textrm{SYM}^n(\mathcal{A}^{(l)}))$
different matrices $A_{\{i_1,\ldots,i_n\}}$ which we can label by 
$\left\{A_k\right\}_{k=1}^d$.
It follows from (\ref{eqWillBeRational}) that these matrices
have integer entries.

They are stored as a list of $2^{ln}\times 2^{ln}$-tables of integers.
Thus, this step of the computation is exact, that is without approximations.
\item[6.] For every $i\in\{1,\ldots,M\}$ and $k\in\{1,\ldots,d\}$, let
\[
   |u_{k,i}\rangle:=A_k|\omega_i\rangle\,\,,
\]
where $|\omega_i\rangle$ denotes the computational basis vector which
is a tensor product of $|0\rangle$'s and $|1\rangle$'s according to
the bits of the string $\omega_i$.
Compute the vectors $|u_{k,i}\rangle$ one after the other. For every
vector that has been computed, check if it can be written as a
linear combination of already computed vectors.
(The corresponding system of linear equations can
be solved exactly, since every vector is given as an array
of integers.) If yes, then discard the new vector
$|u_{k,i}\rangle$, otherwise store it and give it a number.

This way, a set of vectors $\left\{|u_k\rangle\right\}_{k=1}^D$ is
computed. These vectors linearly span the support of the projector
$W_{l,R}^{(ln)}$ given in (\ref{eqJoin2}).
\item[7.] Denote by $\left\{|\phi_i\rangle\right\}_{i=1}^{2^{m-ln}}$
the computational basis vectors of $\hr_{m-ln}$. If $m= l\cdot
2^{3\cdot l}$, then let $\tilde D:=D$, and let
$|x_k\rangle:=|u_k\rangle$. Otherwise, compute
$|u_k\rangle\otimes|\phi_i\rangle$ for every $k\in\{1,\ldots,D\}$
and $i\in\{1,\ldots,2^{m-ln}\}$. The resulting set of vectors
$\left\{|x_k\rangle\right\}_{k=1}^{\tilde D}$ has cardinality
$\tilde D:=D\cdot 2^{m-ln}$.

In both cases, the resulting vectors $|x_k\rangle\in\hr_m$ will span
the support of the projector $Q_{s,\eps}^{(m)}=q_m$.
\item[8.] The set $\left\{|x_k\rangle\right\}_{k=1}^{\tilde D}$ is completed
to linearly span the whole space $\hr_m$. This will be
accomplished as follows:

Consider the sequence of vectors
\[
   (|\tilde x_1\rangle,|\tilde x_2\rangle,\ldots,|\tilde x_{\tilde D+2^m}
\rangle):=   (|x_1\rangle,|x_2\rangle,\ldots,|x_{\tilde D}\rangle,
   |\Phi_1\rangle,|\Phi_2\rangle,\ldots,|\Phi_{2^m}\rangle),
\]
where $\left\{\Phi_k\right\}_{k=1}^{2^m}$ denotes the computational
basis vectors of $\hr_m$. Find the smallest $i$ such that $|\tilde
x_i\rangle$ can be written as a linear combination of $|\tilde
x_1\rangle, |\tilde x_2\rangle, \ldots,|\tilde x_{i-1}\rangle$, and
discard it (this can still be decided exactly, since all the vectors
are given as tables of integers). Repeat this step $\tilde
D$ times until there remain only $2^m$ linearly independent vectors,
namely all the $|x_j\rangle$ and $2^m-\tilde D$ of the
$|\Phi_j\rangle$.
\item[9.] Apply the Gram-Schmidt orthonormalization
procedure to the resulting vectors, to get
an orthonormal basis $\left\{|y_k\rangle\right\}_{k=1}^{2^m}$ of $\hr_m$,
such that
the first $\tilde D$ vectors are a basis for the support of 
$Q_{s,\eps}^{(m)}=q_m$.

Since every vector $|x_j\rangle$ and $|\Phi_j\rangle$ has only
integer entries, all the resulting vectors $|y_k\rangle$ will have
only entries that are (plus or minus) the square root of some
rational number.
\end{itemize}

Up to this point, every calculation was {\em exact} without any
numerical error,
comparable to the way that well-known computer algebra systems work.
The goal of the next steps is to compute an approximate description of the
desired unitary decompression map $U$ and subsequently apply it
to the quantum state $\tilde q$.

According to Section 6 in \cite{BernsteinVazirani}, a {\sl UQTM} is
able to apply a unitary transformation $U$ on some segment of its
tape within an accuracy of $\delta$, if it is supplied with a
complex matrix $\tilde U$ as input which is within operator norm
distance $\frac \delta{2(10\sqrt d)^d}$ of $U$ (here, $d$ denotes
the size of the matrix). Thus, the next task is to compute the number
of digits $N$ that are necessary to guarantee that the output will
be within trace distance $\delta=\frac 1 k$ of $q$.
\begin{itemize}
\item[10.] Read the value of $k$ (which denotes an approximation parameter; the
larger $k$, the more accurate the output of the algorithm will be).
Due to the considerations above and the calculations below, the
necessary number of digits $N$ turns out to be
$N=1+\lceil \log(2k 2^m(10\sqrt {2^m})^{2^m})\rceil$.
Compute this number.

Afterwards, compute the components of all the vectors
$\left\{|y_k\rangle\right\}_{k=1}^{2^m}$ up to $N$ binary digits of
accuracy. (This involves only calculation of the square root of
rational numbers, which can easily be done to any desired accuracy.)

Call the resulting numerically approximated vectors $|\tilde
y_k\rangle$. Write them as columns into an array (a matrix) $\tilde
U:=\left(\tilde y_1,\tilde y_2,\ldots,\tilde y_{2^m}\right)$.

Let $U:=\left(y_1,y_2,\ldots,y_{2^m}\right)$ denote the unitary
matrix with the exact vectors $|y_k\rangle$ as columns. Since $N$
binary digits give an accuracy of $2^{-N}$, it follows that
\[
   \left\vert\tilde U_{i,j}-U_{i,j}\right\vert < 2^{-N} <
   \frac{1/k}{2\cdot 2^m(10\sqrt {2^m})^{2^m}}\,\,.
\]
If two $2^m\times 2^m$-matrices $U$ and $\tilde U$ are $\eps$-close
in their entries, they also must be $2^m\cdot\eps$-close in norm, so
we get
\[
   \|\tilde U-U\|<\frac{1/k}{2(10\sqrt {2^m})^{2^m}}\,\,.
\]
\end{itemize}
So far, every step was purely classical and could have been done on
a classical computer.
Now, the quantum part begins: $\tilde q$ will be touched for the first time.
\begin{itemize}
\item[11.] Compute $\lceil mr\rceil$, which gives the length $\ell(\tilde q)$.
Afterwards, move $\tilde q$ to some free space on the input tape,
and append zeroes, i.e. create the state
\[
   q'\equiv|\psi_0\rangle\langle\psi_0|:=\left(|0\rangle\langle
   0|\right)^{\otimes(m-\ell(\tilde q))}\otimes\tilde q
\]
on some segment of $m$ cells on the input tape.
\item[12.]
Approximately apply the unitary transformation $U$ on the tape segment
that contains the state $q'$.

The machine cannot apply $U$ exactly (since it only knows an
approximation $\tilde U$), and it also cannot apply $\tilde U$
directly (since $\tilde U$ is only approximately unitary, and the
machine can only do unitary transformations). Instead, it will
effectively apply another unitary transformation $V$ which is close
to $\tilde U$ and thus close to $U$, such that
\[
\| V-U \|<\frac 1 k\,\,.
\]

Let $|\psi\rangle:=U|\psi_0\rangle$ be the output that we want to have, and let
$|\phi\rangle:=V|\psi_0\rangle$ be the approximation that is really computed
by the machine. Then,
\[
\|\, |\phi\rangle-|\psi\rangle\| <\frac 1 k\,\,.
\]
A simple calculation proves that the trace distance must then also be small:
\[
\||\phi\rangle\langle\phi|-|\psi\rangle\langle\psi|\|_{\rm Tr}<\frac 1 k\,\,.
\]

\item[14.] Move $q:=|\phi\rangle\langle\phi|$ to the output tape and halt.
\end{itemize}
\subsubsection{Proof of Proposition~\ref{PropUpperBound}}
We have to give a precise definition how the parameters $(m,r,\tilde
q)$ are encoded into a single qubit string $\sigma$. (According to
the definition of $QC^{\searrow 0}$, the parameter $k$ is not a part
of $\sigma$, but is given as a second parameter. See 
Definitions \ref{DefEncoding} and \ref{defQK} for details.)

We choose to encode $m$ by giving $\lfloor\log m\rfloor$ 1's, followed
by one 0, followed by the  $\lfloor\log m\rfloor +1$ binary digits of $m$.
Let $|M\rangle\langle M|$ denote the corresponding projector in the
computational basis.

The parameter $r$ can be encoded in any way, since it does not
depend on $m$. The only constraint is that the description must be
self-delimiting, i.e. it must be clear and decidable at what
position the description for $r$ starts and ends. The descriptions
will also be given by a computational basis vector (or rather the
corresponding projector) $|R\rangle\langle R|$.

The descriptions are then stuck together, and the input 
$\sigma(\tilde q)$ is given by
\[
   \sigma(\tilde q):=|M\rangle\langle M|\otimes|R\rangle\langle R|\otimes
   \tilde q\,\,.
\]
If $m$ is large enough such that (\ref{eqLogTrQm}) is fulfilled, it
follows that
$\ell(\sigma(\tilde q))=2\lfloor\log m\rfloor + 2 + c + \lceil mr\rceil$,
where $c\in\N$ is some constant which depends on $r$, but not on $m$.

It is clear that this qubit string can be fed into the reference
{\sl UQTM} $\mathfrak{U}$ together with a description of the
algorithm $\mathfrak{A}$ of fixed length $c'$ which depends on $r$, 
but not on $m$. This will give a
qubit string $\sigma_{\mathfrak U}(\tilde q)$ of length
\begin{eqnarray}
   \ell(\sigma_{\mathfrak U}(\tilde q))&=&2\lfloor\log m\rfloor
   +2+c+\lceil mr\rceil+c'\nonumber\\
   &\leq& 2 \log m + m\left(s+\frac 1 2 \delta\right)+c''\,\,,
   \label{eqLength}
\end{eqnarray}
where $c''$ is again a constant which depends on $r$, but not on $m$.
Recall the matrix $U$ constructed in step 11 of our algorithm
$\mathfrak A$, which rotates (decompresses) a compressed
(short) qubit string $\tilde q$ back into the typical subspace.
Conversely, for every one-dimensional projector $q\leq q_m$, where
$q_m=Q_{s,\eps}^{(m)}$ was defined in (\ref{eqTypicalProjector}), let
$\tilde q\in\hr_{\lceil mr\rceil}$ be the projector given by
$\left(|0\rangle\langle 0|\right)^{\otimes(m-\lceil
mr\rceil)}\otimes \tilde q =U^* q U$. Then, since $\mathfrak A$ has
been constructed such that
\[
   \|\mathfrak{U}(\sigma_{\mathfrak U}(\tilde q),k)-q\|_{\rm Tr}
   < \frac 1 k\qquad\mbox{ for every }k\in\N\,\,,
\]
it follows from (\ref{eqLength}) that
\[
   \frac 1 m QC^{\searrow 0}(q)\leq 2 \frac{\log m} m + s+
   \frac 1 2 \delta +\frac{c''}m\,\,.
\]
If $m$ is large enough, Equation~(\ref{UpperZero}) follows.

Now we continue by proving Equation~(\ref{UpperDelta}).
Let $k:=\lceil\frac 1 {2\delta}\rceil$. Then, we have
for every one-dimensional projector $q\leq q_m$ and $m$ large enough
\begin{eqnarray}
   \frac 1 m QC^{2\delta}(q)&\leq& \frac 1 m QC^{1/k}(q)
   \leq \frac 1 m QC^{\searrow 0}(q)
   +\frac{2\lfloor\log k\rfloor+2} m\nonumber\\
   &<&s+\delta  +\frac{2\log k +2} m<s+2\delta\,\,,
   \label{eqOmegaEquation}
\end{eqnarray}
where the first inequality follows from the obvious monotonicity property
$\delta\geq \eps \Rightarrow QC^\delta\leq QC^\eps$, the second one is
by Lemma~\ref{LemRelation},
and the third estimate is due to Equation~(\ref{UpperZero}).
\qed

\textit{Proof of the Main Theorem~\ref{TheQBrudno}.} Let $\tilde
q_m(\delta)$ be the $\Psi$-typical projector sequence given in 
Proposition~\ref{PropUpperBound}, i.e.
the complexities $\frac 1 m QC^{\searrow 0}$ and $\frac 1 m QC^\delta$ of every
one-dimensional projector $q\leq \tilde q_m(\delta)$ are upper bounded 
by $s+\delta$.
Due to Corollary~\ref{Cor1}, there exists another sequence of
$\Psi$-typical projectors
$p_m(\delta)\leq \tilde q_m(\delta)$ such that additionally, 
$\frac 1 m QC^\delta(q)>s-\delta(2+\delta)s$
is satisfied for $q\leq p_m(\delta)$. From Corollary~\ref{Cor2}, 
we can further deduce that there is another sequence of 
$\Psi$-typical projectors $q_m(\delta)\leq p_m(\delta)$
such that also $\frac 1 m QC^{\searrow 0}(q)>s-\delta$ holds. Finally, the optimality assertion is a direct consequence of the Quantum Counting Argument, Lemma \ref{CountingArgument}, combined with Theorem \ref{QAEP}. \qed

\section{Summary and Perspectives}
Classical algorithmic complexity theory as initiated by Kolmogorov,
Chaitin and Solomonoff aimed at giving firm mathematical ground to
the intuitive notion of randomness. The idea is that random objects
cannot have short descriptions. Such an approach is on the
one hand equivalent to Martin-L\"of's which is based on the notion
of {\em typicalness}~\cite{USS}, and is on the other hand
intimately connected with the notion of entropy. The latter relation
is best exemplified in the case of longer and longer strings: by
taking the ratio of the complexity with respect to the number of
bits, one gets a \textit{complexity per symbol} which a theorem of
Brudno shows to be equal to the \textit{entropy per symbol} of
almost all sequences emitted by ergodic sources.

The fast development of quantum information and computation, with
the formalization of the concept of {\sl UQTMs}, quite naturally
brought with itself the need of extending the notion of algorithmic
complexity to the quantum setting. Within such a broader context,
the ultimate goal is again a mathematical theory of the randomness
of quantum objects. There are two possible
algorithmic descriptions of qubit strings: either by
means of bit-programs or of qubit-programs.
In this work, we have considered a qubit-based
\textit{quantum algorithmic complexity}, namely constructed in terms
of quantum descriptions of quantum objects.

The main result of this paper is an extension of Brudno's theorem
to the quantum setting,
though in a slightly weaker form which is due to the
absence of a natural concatenation of qubits. The quantum Brudno's
relation proved in this paper is not a pointwise relation as in the
classical case, rather a kind of convergence in probability which
connects the {\it quantum complexity per qubit} with the von Neumann
entropy rate of quantum ergodic sources. Possible strengthening of
this relation following the strategy which permits the formulation
of a quantum Breiman theorem starting from the quantum Shannon-McMillan
noiseless coding theorem~\cite{BKSSS} will be the matter of future
investigations.

In order to assert that this choice of quantum complexity as a
formalization of ''quantum randomness" is as good as its classical
counterpart in relation to ''classical randomness'', one ought to
compare it with the other proposals that have been put forward: not
only with the quantum complexity based on classical descriptions of
quantum objects~\cite{Vitanyi}, but also with the one based on the
notion of \textit{universal density matrices}~\cite{Gacs}.

In relation to Vitanyi's approach, the comparison essentially boils
down to understanding whether a classical description of qubit
strings requires more classical bits than $s$ qubits per Hilbert
space dimension. An indication that this is likely to be the case
may be related to the existence of entangled states.

In relation to Gacs' approach, the clue is provided by the possible
formulation of ''quantum Martin-L\"of'' tests in terms of measurement
processes projecting onto low-probability subspaces, the quantum
counterparts of classical untypical sets.

One cannot however expect classical-like equivalences among the
various definitions. It is indeed a likely consequence of the very
structure of quantum theory that a same classical notion may be
extended in different inequivalent ways, all of them reflecting a
specific aspect of that structure. This fact is most clearly seen in
the case of quantum dynamical entropies (compare for
instance~\cite{AN}) where one definition can capture dynamical
features which are precluded to another. Therefore, it is possible
that there may exist different, equally suitable notions of "quantum
randomness", each one of them reflecting a different facet of it.

\vskip 0.5cm

\begin{acknowledgements}
We would like to thank our colleagues Ruedi Seiler and
Igor Bjelakovi\'c for their constant encouragement and for
helpful discussions and suggestions.

This work was supported by the DFG via the
project ``Entropie, Geometrie und Kodierung gro\ss er
Quanten-Informationssysteme'' and
the DFG-Forschergruppe ``Stochastische Analysis und gro\ss e
Abweichungen'' at the University of Bielefeld.
\end{acknowledgements}


\begin{thebibliography}{99}

\bibitem{ADMDH}
L.M. Adleman, J. Demarrais and M. A. Huang, "Quantum Computability", {\em SIAM J. Comput.} {\bf
26} 1524-1540 (1997)

\bibitem{AlYak} V.M. Alekseev, M.V. Yakobson, "Symbolic Dynamics and Hyperbolic Dynamic Systems", {\it Phys. Rep.} {\bf
75} 287 (1981).

\bibitem{AN}
R. Alicki and H. Narnhofer, "Comparison of dynamical entropies for the
noncommutative shifts", {\em Lett. Math. Phys.} {\bf 33} 241-247
(1995)

\bibitem{BernsteinVazirani} E. Bernstein, U. Vazirani, "Quantum Complexity Theory", {\em
SIAM Journal on Computing} {\bf 26} 1411-1473 (1997)

\bibitem{Berthiaume}
A. Berthiaume, W. Van Dam and S. Laplante, "Quantum Kolmogorov complexity", {\em J. Comput, System
Sci.} {\bf 63} 201-221 (2001)

\bibitem{Billingsley}
P. Billingsley, "Ergodic Theory and Information", {\em Wiley Series
in Probability and Mathematical Statistics, John Wiley \& Sons, New
York 1965}

\bibitem{QSMPaper}
I. Bjelakovi\'c, T. Kr\"uger, Ra. Siegmund-Schultze, A. Szko\l a, "The Shannon-
McMillan theorem for ergodic quantum lattice systems",
{\em Invent. Math.} {\bf 155} 203-222 (2004)

\bibitem{BKSSS}
I. Bjelakovi\'c, T. Kr\"uger, Ra. Siegmund-Schultze and A. Szko\l a,
 "Chained Typical Subspaces- a Quantum Version of Breiman's
Theorem", \texttt{quant-ph/0301177}

\bibitem{BjelSzk}
I. Bjelakovi\'c and A. Szko\l a, "The Data Compression Theorem for
Ergodic Quantum Information Sources", {\em Quant. Inform. Proc.} {\bf 4 No. 1} 49-63 (2005)

\bibitem{Brudno}
A. A. Brudno, "Entropy and the complexity of the trajectories of a dynamical
system", {\em Trans. Moscow Math. Soc.} {\bf 2} 127-151 (1983)

\bibitem{Chaitin} G. J. Chaitin, "On the Length of Programs for Computing Binary Sequences", {\em J. Assoc. Comp. Mach.} {\bf 13} 547-569 (1966)

\bibitem{CoverThomas}
T. M. Cover, J. A. Thomas, ''Elements of Information Theory'',
{\em Wiley Series in Telecommunications, John Wiley \& Sons, New York 1991}

\bibitem{Deutsch}
D. Deutsch, Quantum theory, the Church-Turing principle and the universal quantum
computer, {\em Proc. R. Soc. Lond.}, {\bf A400} (1985)

\bibitem{Feynman}
R. Feynman, Simulating physics with computers, {\em International Journal
of Theoretical Physics}, {\bf 21}(1982) pp. 467-488

\bibitem{Gacs}
P. G\'acs, "Quantum algorithmic entropy", {\em J. Phys. A: Math. Gen.} {\bf 34} 6859-6880 (2001)

\bibitem{Gr}
J. Gruska, "Quantum Computing", {\em McGraw--Hill, London 1999}

\bibitem{Hiai} F. Hiai, D. Petz, "The Proper Formula for Relative Entropy and its Asymptotics in Quantum Probability", {\em Commun. Math. Phys.} {\bf 143} 99-114 (1991)

\bibitem{holevo} A. S. Holevo, ''Statistical Structure of Quantum Theory'',
{\em Springer Lecture Notes} {\bf 67} (2001)

\bibitem{Jozsa}
R. Jozsa and B. Schumacher, "A new proof of the quantum noiseless coding theorem", {\em J. Mod. Optics} {\bf 41} 2343-2349
(1994)

\bibitem{JHHH}
R. Jozsa, M. Horodecki, P. Horodecki and R. Horodecki, "Universal Quantum Information Compression" {\em Phys.
Rev. Lett.} {\bf 81} 1714-1717 (1998)

\bibitem{KaltchenkoYang}
A. Kaltchenko and E. H. Yang, "Universal compression of ergodic quantum
sources", {\em Quantum Information and
Computation} {\bf 3, No. 4} 359-375 (2003)

\bibitem{Keller}
G. Keller, "Wahrscheinlichkeitstheorie", {\em Lecture Notes,
Universit\"at Erlangen-N\"urnberg} (2003)

\bibitem{Kieffer} J. Kieffer, "A unified approach to weak universal source coding", {\em IEEE Trans. Inform. Theory} {\bf
24 No. 6} 674-682 (1978)

\bibitem{Kolmogorov65} A. N. Kolmogorov, Three Approaches to the Quantitative Definition on Information, {\em Problems of Information Transmission} {\bf 1} 4-7 (1965)

\bibitem{Kolmogorov68} A. N. Kolmogorov, "Logical Basis for Information Theory and Probability Theory", {\em IEEE Trans. Inform. Theory} {\bf 14}, 662-664 (1968)

\bibitem{Vitanyibook}
M. Li and P. Vitanyi, "An Introduction to Kolmogorov Complexity
and Its Applications", {\em Springer Verlag 1997}

\bibitem{Briegel1}
C. Mora and H. J. Briegel, "Algorithmic complexity of quantum
states", \texttt{quant-ph/0412172}

\bibitem{Briegel2}
C. Mora and H. J. Briegel, ''Algorithmic complexity and entanglement of quantum states'',
\texttt{quant-ph/0505200}

\bibitem{NielsenChuang}
M. A. Nielsen, I. L. Chuang, "Quantum Computation and Quantum
Information", {\em Cambridge University Press, Cambridge, 2000}

\bibitem{Ozawa}
M. Ozawa and H. Nishumura, ''Local Transition Functions of Quantum
Turing Machines'', {\em Theoret. Informatics and Appl.} {\bf 34} (2000) 379-402

\bibitem{paulsen}
V. Paulsen, ''Completely Bounded Maps and Operator Algebras'',
{\em Cambridge Studies in Advanced Mathematics} {\bf 78} (2002)

\bibitem{Perdrix}
S. Perdrix, P. Jorrand, "Measurement-Based Quantum Turing
Machines and their Universality", \texttt{quant-ph/0404146}

\bibitem{MosonyiPetz}
D. Petz and M. Mosonyi, "Stationary quantum source coding", {\em J. Math. Phys.} {\bf 42} 4857-4864 (2001)

\bibitem{Shor}
P. Shor, ''Algorithms for quantum computation: Discrete log and factoring'', {\em Proceedings
of the 35th Annual IEEE Symposium on Foundations of Computer Science} (1994)

\bibitem{Segre}
G. Segre, ''Physical Complexity of Classical and Quantum Objects and Their
Dynamical Evolution From an Information-Theoretic Viewpoint'', {\em Int. J. Th. Phys.}
{\bf 43} 1371-1395 (2004)

\bibitem{Solomonoff} R. J. Solomonoff, "A Formal Theory of Inductive Inference", {\em Inform. Contr.} {\bf 7} 1-22, 224-254 (1964)

\bibitem{Sow}
D. M. Sow, A. Eleftheriadis, "Complexity distortion theory", {\em IEEE Trans. Inform. Theory},
{\bf IT-49}, 604-608 (2003)

\bibitem{Svozil}
K. Svozil, ''Randomness and Undecidability in Physics'', World Scientific (1993)

\bibitem{USS}
V.A. Uspenskii, A.L. Semenov and A.Kh. Shen, "Can an individual sequence of zeros and ones
be random", {\em Uspekhi Mat.
Nauk.} {\bf 45/1} 105-162 (1990)

\bibitem{Vitanyi}
P. Vit\'anyi, "Quantum Kolmogorov complexity based on classical descriptions", {\em IEEE Trans. Inform. Theory} {\bf 47/6} 2464-2479 (2001)

\bibitem{Whi}
H. White, "Algorithmic complexity of points in a dynamical system", {\it Erg. Th. Dyn. Sys.} {\bf 13} 807 (1993)

\bibitem{Ziv} J. Ziv, "Coding of sources with unknown statistics--I: Probability of encoding error", {\em IEEE Trans. Inform. Theory} {\bf 18} 384-389 (1972)

\bibitem{Zvonkin}
A. K. Zvonkin, L. A. Levin, "The complexity of finite objects and the development of the
concepts of information and randomness by means of the theory of algorithms", {\em Russian Mathematical
Surveys} {\bf 25 No. 6} 83-124 (1970)

\end{thebibliography}
\end{document}